\documentclass[aps,twocolumn,showpacs,prb,floatfix]{revtex4-1}

\usepackage{color}
\usepackage{bm}
\usepackage[dvipdfmx]{graphicx}
\usepackage{amsmath,amssymb}
\usepackage{newtxtext,newtxmath}
\usepackage{accents}
\usepackage{float}
\usepackage{wrapfig}
\usepackage{layout}
\usepackage{braket}
\let\%\relax 
\DeclareFontFamily{OT1}{pxr}{}
\DeclareFontShape{OT1}{pxr}{m}{n}{<->pxr}{}
\DeclareSymbolFont{letA}{OT1}{pxr}{m}{n}
\DeclareMathSymbol{\%}{0}{letA}{`\%}
\usepackage{scalerel,stackengine}
\stackMath
\newcommand\reallywidecheck[1]{%
\savestack{\tmpbox}{\stretchto{%
  \scaleto{%
    \scalerel*[\widthof{\ensuremath{#1}}]{\kern-.6pt\bigwedge\kern-.6pt}%
    {\rule[-\textheight/2]{1ex}{\textheight}}
  }{\textheight}%
}{0.5ex}}%
\stackon[1pt]{#1}{\scalebox{-1}{\tmpbox}}%
}
\newcommand{\tPt}{${}^{3}P_{2}$ }
\newcommand{\Car}{\mathrm{Car}}
\newcommand{\AM}{\mathrm{AM}}
\newcommand{\Cyl}{\mathrm{Cyl}}
\newcommand{\ii}{\mathrm{i}}
\newcommand{\F}{\mathrm{F}}
\newcommand{\dd}{\mathrm{d}}

\newcommand{\BdG}{\mathrm{BdG}}

\newcommand{\n}{\mathrm{n}}

\newcommand{\Tr}{\mathrm{Tr}}

\begin{document}
\preprint{aps/}
\title{Microscopic description of axisymmetric vortices in $^{3}P_{2}$ superfluids}

\author{Yusuke Masaki}
\email{masaki@cmpt.phys.tohoku.ac.jp}
\affiliation{Research and Education Center for Natural Sciences, Keio University, Hiyoshi 4-1-1,
Yokohama, Kanagawa 223-8521, }
\affiliation{Department of Physics, Tohoku University, Sendai, Miyagi 980-8578, Japan}
\author{Takeshi Mizushima}
\affiliation{Department of Materials Engineering Science, Osaka University, Toyonaka, Osaka 560-8531, Japan}
\author{Muneto Nitta}
\affiliation{Department of Physics, Keio University, Hiyoshi 4-1-1,}
\affiliation{Research and Education Center for Natural Sciences, Keio University, Hiyoshi 4-1-1,
Yokohama, Kanagawa 223-8521, Japan}
\date{\today}
\begin{abstract}
We study quantized vortices in \tPt superfluids using a microscopic theory for the first time. 
The theory is based on the Eilenberger equation to determine the order parameters and the Bogoliubov-de Gennes (BdG) equation to obtain the eigenenergies and the core magnetization.
Within axisymmetric vortex configurations, 
we find several stable and metastable vortex configurations which depend on the strength of a magnetic field, 
similar to a $v$ vortex and $o$ vortex 
in $^3$He superfluids. 
We demonstrate that 
the $o$ vortex is the most stable axisymmetric vortex in the presence of a strong magnetic field, and we find two zero-energy Majorana fermion bound states in the $o$-vortex core.
We show that the profiles of the core magnetization calculated using the BdG equation are drastically different from those calculated using only the order parameter profiles known before.
\end{abstract}

\maketitle

\section{Introduction}
Superfluidity and superconductivity are two of the most extraordinary states of matter. They are realized in materials or gases at low temperatures, 
such as metals, liquid heliums, and 
Bose-Einstein condensates of ultracold atomic gases.
Apart from such situations realized in laboratory experiments, 
neutron stars offer much larger, stellar scale candidates of 
superfluidity and superconductivity \cite{Migdal:1960,Reichely1969,Baym1969}.  
The neutron density in the interior of neutron stars ranges from $10^{4}\mathrm{g/cm^{3}}$ to $10^{16}\mathrm{g/cm^{3}}$ and 
forms a hierarchical structure consisting of crusts and cores. 
Neutrons in the inner crust and outer core drip from nuclei and become
a neutron fluid. The superfluidity is important to such a high density region because the temperature $T\sim 10^{6}\mathrm{K}$ is much lower than the Fermi temperature $T_{\F}\sim 10^{10} \mathrm{K}$ and the critical temperature $T_{\mathrm{c}}\sim10^{8}\mathrm{K}$. 
The presence of Cooper pairs successfully describes rapid coolings 
\cite{Yakovlev:1999sk} of neutron stars 
and slow relaxations \cite{Reichely1969}
after {\it pulser glitches}, i.e. phenomena in which angular momentum of neutron stars increases suddenly 
(see Refs.~[\onlinecite{Chamel2017,Haskell:2017lkl,Sedrakian2019}] 
for a recent review). 

A type of the Cooper instability depends on the density. 
The attractive interaction is governed by the ${}^{1}S_{0}$ channel in the inner crust at lower density \cite{Migdal:1960},
while in the outer core at higher density it becomes repulsive and the \tPt channel is dominant 
\cite{Tabakin:1968zz,Hoffberg:1970vqj,Tamagaki1970,Takatsuka1971,Takatsuka1972,Fujita1972,Richardson:1972xn,Amundsen:1984qc,Takatsuka:1992ga,Baldo:1992kzz,Elgaroy:1996hp,Khodel:1998hn,Baldo:1998ca,Khodel:2000qw,Zverev:2003ak,Maurizio:2014qsa,Bogner:2009bt,Srinivas:2016kir}.
The \tPt superfluid is the state in which the angular momenta of orbital ($L=1$) and spin ($S=1$) are aligned.  
The Ginzburg--Landau (GL) equation was obtained for  \tPt superfluids 
\cite{Richardson:1972xn,Fujita1972,Sauls:1978lna,Muzikar:1980as,Sauls:1982ie} 
and the ground state was found to be a nematic phase in the weak coupling regime  
\cite{Sauls:1978lna}.  
The nematic phase has a continuous degeneracy \footnote{
In spin-2 Bose-Einstein condensates, 
this is known to be related to the existence of quasi-Nambu-Goldstone modes 
\cite{Uchino:2010pf}.
} which is lifted by either magnetic field or sixth order terms in the GL free energy, 
and  
the uniaxial nematic (UN) phase is favored at zero temperature 
while $D_{2}$-biaxial nematic (BN) and $D_{4}$-BN phases are favored for finite and strong magnetic fields, respectively \cite{Masuda:2015jka}, relevant for magnetars. 
Low-energy excitations in \tPt superfluids 
affect the cooling process by neutrino emission~\cite{Bedaque:2003wj,Bedaque:2012bs,Bedaque:2013fja,Bedaque:2014zta,Leinson:2009nu,Leinson:2010yf,Leinson:2010pk,Leinson:2010ru,Leinson:2011jr,Leinson:2012pn,Leinson:2013si,Leinson:2014cja}.
The rapid cooling due to \tPt superfluids was studied for Cassiopeia A~\cite{Heinke2010,Shternin2011,Page:2010aw}, but a direct proof of the existence of the \tPt superfluidity is yet elusive. 
The \tPt superfluidity is more relevant for magnetars, in which the strength of the magnetic field reaches about $10^{15}$ G at the surface and possibly about $10^{18}$ G in the inside, 
because the tolerance of the spin-triplet pairing in the \tPt superfluidity is robust
against the strong magnetic fields,
in contrast to the spin-singlet pairing in the $^{1}S_{0}$ superfluidity fragile due to the Zeeman effect.
The impact of strong magnetic fields on \tPt superfluid phases was studied in the GL equation \cite{Yasui:2018tcr}.
 The GL equation was also used for finding new exotic topological structures such as surface topological defects (boojums) \cite{Yasui:2019pgb} and domain walls \cite{Yasui:2019vci}.
 
Since neutron stars rotate very rapidly, there exist a huge amount of quantized vortices. 
These vortices may play an important role on the glitches \cite{Anderson:1975zze}; One of scenarios of the glitches is described by unpinning of a large amount of vortices which transfers the angular momentum from the superfluid to the nonsuperfluid crust.
Vortices in \tPt superfluids have been studied using the GL equation 
\cite{Richardson:1972xn,Fujita1972,Sauls:1978lna,Muzikar:1980as,
Sauls:1982ie,Masuda:2015jka,Masuda:2016vak,Chatterjee:2016gpm}.
Because of a discrete symmetry of the condensates, we expect rich structures of vortices, such as fractional vortices and non-Abelian vortices \cite{Masuda:2016vak}. 
In addition to such nontrivial topology, spontaneous  magnetization 
\cite{Sauls:1982ie,Masuda:2015jka,Masuda:2016vak,Chatterjee:2016gpm} 
is a crucial issue to explain the above phenomena.  For further study of vortices beyond the GL equation, it  is more important to take account of the fermion degrees of freedom because fermions form vortex bound states. Therefore, in this paper, we formulate a microscopic theory and calculate single vortex states with or without a magnetic field.

The microscopic model of \tPt superfluids was constructed long back by Richardson \cite{Richardson:1972xn} and 
Tamagaki and Takatsuka \cite{Tamagaki1970,Takatsuka1971,Takatsuka1972}, 
but the first direct calculation was done recently 
\cite{Mizushima:2016fbn},
which clarifies that nematic states of \tPt superfluid is a topological superfluid with time reversal symmetry (a class DIII in the classification of topological insulators and superconductors), allowing gapless Majorana fermion on the boundary. 
The existence of such fermion bound states is a noticeable character of the topological states.  
The microscopic theory also offers
the phase diagram for the magnetic field and temperature 
and elucidates that there is a tricritical point on the phase boundary between $D_{2}$-BN and $D_{4}$-BN states \cite{Mizushima:2016fbn,Mizushima:2019spl}, 
which was later confirmed in the GL free energy up to the eighth order \cite{Yasui2019}. 
Moreover, cyclic and ferromagnetic phases are possible for total spin-2 condensates 
\cite{Mermin:1974zz}, 
and these states have been shown to be Weyl semimetals for  \tPt superfluids
\cite{Mizushima:2016fbn, Mizushima:2017pma}, 
having gapless Weyl fermions in the bulk 
which may be relevant for cooling of neutron stars.

In general, for topological superconductors and superfluids, Majorana fermions may exist in the vortex core as well as their boundary. 
The topological aspect of fermion degrees of freedom emergent in vortices of \tPt superfluids has not yet been uncovered. In the superfluid $^3$He, which is a prototype of spin-triplet $p$-wave superfluid, all possible vortices are classified in terms of discrete symmetries preserved in vortex states~\cite{Salomaa1987}. 
The $o$-vortex and $v$-vortex states are the local minima of the thermodynamic potential in the superfluid $^3$He-B under rotation, where the former (latter) preserves (breaks) the magnetic $\pi$ rotation symmetry called the $P_{3}$ symmetry~\cite{salomaaPRL83,Passvogel1984,thunebergPRL86,kasamatsuPRB19,Regan2019}. \tPt superfluids with vortices are categorized into the class D of the topological periodic table, and remaining discrete symmetry plays a critical role on the topological protection of the zero energy vortex-bound states~\cite{Teo2010,shiozakiPRB14}. The $o$ vortex is the most symmetric vortex with spin-degenerated zero modes, which are protected by the $P_3$ symmetry~\cite{Tsutsumi2015,mizushimaJPSJ16}. In contrast, the $v$ vortex which spontaneously breaks the $P_3$ symmetry has no topologically protected zero modes. In the superfluid $^3$He-B under rotation, the $o$-vortex state is not thermodynamically stable against  axisymmetric and nonaxisymmetric $v$-vortex states with no zero modes. Therefore, it is an important issue to study if a vortex with zero energy Majorana fermions is possible in \tPt superfluids. 

The existence of fermion degrees of freedom is also 
important at the macroscopic level, as pointed out by Jones \cite{Jones2009}. 
As is known in superfluid and superconducting systems, fermion bound states in the vortex core seriously affect vortex dynamics through the spectral flow force~\cite{kopninPRB91,Kopnin_2002,Volovik2014}.
The vortex dynamics is a key role in interpreting a gradual decrease of angular momentum of a neutron star and its glitch.
Understanding the self-consistent structure of vortices and the topological protection of vortex-bound states in \tPt superfluids may open a door to the issues in neutron stars.

In this paper, 
motivated by these earlier works, we investigate vortex states and fermion bound states in the vortex core in \tPt superfluids using the microscopic theory. The theory is 
based on the Eilenberger equation to determine the order parameters and the Bogoliubov-de Gennes (BdG) equation to obtain the eigenenergies and the core magnetization.
We investigate several stable and metastable vortices and discuss their stability with respect to their free energies in the presence of a magnetic field. 
We also clarify if Majorana fermions exist or not in the vortex core using the BdG equation, and calculate spin densities around the vortex core. 
We find, in the presence of axisymmetry, an $o$ vortex is stable for strong magnetic field and allows spin-degenerate (two) zero-energy Majorana fermions in its core.
This finding may be important to comprehension of the cooling rate and the changes of rotating rate of neutron stars.
We also show that the profiles of the core magnetization calculated using the BdG equation are drastically different from those calculated using only the order parameter profiles 
in the GL theory \cite{Sauls:1982ie,Masuda:2015jka,Masuda:2016vak,Chatterjee:2016gpm}.

The remaining part of this paper consists as follows. In Sect.~\ref{sec:model}, we explain the microscopic equations of the \tPt superfluid and its axisymmetric condition. In Sect.~\ref{sec:results}, we show the numerical results: We seek for order parameter profiles and their free energy densities on the basis of the quasiclassical scheme. We also obtain the eigenenergies of fermion excitations and core magnetizations which consist of order parameter modulations and fermion bound states using the BdG equation. In Sec.~\ref{sec:summary}, we provide a summery and brief discussion about nonaxisymmetric vortices. In Appendix~\ref{sec:RepCooperPair}, we summarize basis sets of the order parameter and their matrix representations. In Appendices~\ref{sec:AM} and \ref{sec:Rotation}, we review the angular momentum operators of Cooper pairs and the rotation of a basis set, respectively. In Appendix~\ref{sec:ldos-ev}, we compare the fermion excitation spectra obtained by solving the BdG equation and those in the quasiclassical theory.

Throughout the paper, we specify the orthonormal spatial and spin directions by 1,2, and 3, and use the notations $\hat{\cdot}$, and $\check{\cdot}$ for a 2 by 2 and a 4 by 4 matrix, respectively. Particularly, $\hat{\sigma}_{\alpha=1,2,3}$ is the $\alpha$-th component of Pauli matrices. We also set $\hbar=k_{\rm B}=1$.

\section{Model and  Method} \label{sec:model}
\subsection{Gor'kov equation}
First we introduce the microscopic Hamiltonian with a zero range, and attractive \tPt force between neutrons\cite{Richardson:1972xn}:
\begin{align}
	H &= H_{1} + H_{2}, \\
	H_{1} &= \int \dd \bm{r}\sum_{\sigma,\sigma^{\prime}=\uparrow,\downarrow}
	\psi_{\sigma}^{\dagger}(\bm{r})\left(
	h_{0}(\bm{r})\delta_{\sigma,\sigma^{\prime}}+ U_{\sigma\sigma^{\prime}}(\bm{r})\right)\psi_{\sigma^{\prime}}(\bm{r}),\\
	H_{2} &= -\int \dd \bm{r} \sum_{\alpha\beta=1,2,3}\dfrac{g}{2}T_{\alpha\beta}^{\dagger}(\bm{r})T_{\alpha\beta}(\bm{r}). \label{eq-hamiltonian}\end{align}
The second line consists of the kinetic energy $h_{0}(\bm{r}) =(-\frac{\nabla^{2}}{2m} -\mu)$ with a chemical potential $\mu$ and the Zeeman energy $\hat{U}(\bm{r}) = -V_{\mathrm{Z}}\hat{\sigma}_{3}$ of the magnetic field parallel to the third direction. 
In the third line,  the interaction strength $g$ is positive, and $T^{\dagger}(T)$ is a pair creation (annihilation) operator defined by
\begin{align}
T_{\alpha\beta}^{\dagger}(\bm{r}) &= 
\sum_{\sigma\sigma^{\prime}}\psi_{\sigma}^{\dagger}(\bm{r})\left[t_{\alpha\beta,\sigma\sigma^{\prime}}^{*}(\ii\bar{\bm{\nabla}})\psi_{\sigma^{\prime}}^{\dagger}(\bm{r})\right],\\
T_{\alpha\beta}(\bm{r}) &= 
\sum_{\sigma\sigma^{\prime}}\left[t_{\alpha\beta,\sigma\sigma^{\prime}}(-\ii\bar{\bm{\nabla}})\psi_{\sigma^{\prime}}(\bm{r})\right]\psi_{\sigma}(\bm{r}),
\end{align}
where we use the dimensionless notation $\bar{\bm{\nabla}} \equiv k_{\F}^{-1}\bm{\nabla}$. A 2 by 2 matrix in spin space, $\hat{t}_{\alpha\beta}$, is defined for $\alpha, \beta= 1,2,3$ as
\begin{align}
\hat{t}_{\alpha\beta}(-\ii \bar{\bm{\nabla}}) &=\ii\hat{\sigma}_{2} \left\{\dfrac{1}{2\sqrt{2}}\left[\hat{\sigma}_{\alpha}(-\ii\bar{\nabla}_{\beta}) \right.\right.\nonumber \\
&\left.\left.+\hat{\sigma}_{\beta}(-\ii\bar{\nabla}_{\alpha})\right] -\dfrac{1}{3\sqrt{2}}
\delta_{\alpha\beta}\hat{\bm{\sigma}}
\cdot(-\ii\bar{\bm{\nabla}})\right\}.
\end{align}
We see that $[\hat{t}_{\alpha\beta}(-\ii\bar{\bm{\nabla}})]^{*} =  \hat{t}_{\alpha\beta}^{*}(\ii\bar{\bm{\nabla}})$ and that $\hat{t}_{\alpha\beta}$ is symmetric and traceless regarding the subscripts, namely, $T_{\alpha\beta} = T_{\beta\alpha}$. We also find that $t_{\alpha\beta,\sigma\sigma^{\prime}}=t_{\alpha\beta,\sigma^{\prime}\sigma}$. We derive the Gor'kov equation for Green's functions defined by
\begin{align}
\check{G}  (\bm{r}_{1},\bm{r}_{2};\ii\omega_{n})&=
\begin{bmatrix}
\hat{G}(\bm{r}_{1},\bm{r}_{2};\ii\omega_{n}) & \hat{F}(\bm{r}_{1},\bm{r}_{2};\ii\omega_{n})\\
-\hat{\bar{F}}(\bm{r}_{1},\bm{r}_{2};\ii\omega_{n}) & \hat{\bar{G}}(\bm{r}_{1},\bm{r}_{2};\ii\omega_{n})
\end{bmatrix} \nonumber \\
&=\int_{0}^{\beta} \dd \tau e^{\ii \omega_{n} \tau}\check{\tau}_{3}\braket{\mathcal{T}_{\tau}\vec{\Psi}(\bm{r}_{1},\tau)\vec{\Psi}^{\dagger}(\bm{r}_{2})}.
\end{align}
Here $\mathcal{T}_{\tau}$ is the time-ordering operator on the imaginary axis. We have defined $\psi_{\sigma}(\bm{r},\tau) = e^{H\tau}\psi_{\sigma}(\bm{r}) e^{-H\tau}$ and 
$\psi_{\sigma}^{\dagger}(\bm{r},\tau) = e^{H\tau}\psi_{\sigma}^{\dagger}(\bm{r})e^{-H\tau}$ and the Nambu spinor 
$\vec{\Psi}(\bm{r},\tau) \!=\! {}^t (
\psi_{\uparrow}(\bm{r},\tau),
\psi_{\downarrow}(\bm{r},\tau),
\psi_{\uparrow}^{\dagger}(\bm{r},\tau),
\psi_{\downarrow}^{\dagger}(\bm{r},\tau)
)$. 
We perform the Hartree--Fock--Bogoliubov approximation for $H_{2}$ while neglecting its contribution to the one body potential:
$
\braket{T_{\tau}T_{\alpha\beta}(\bm{r}_{1},\tau)\psi_{\sigma^{\prime\prime}}^{\dagger}(\bm{r}_{1},\tau)\psi_{\sigma^{\prime}}^{\dagger}(\bm{r}_{2})}\approx
\braket{T_{\alpha\beta}(\bm{r}_{1},\tau)}\braket{T_{\tau}\psi_{\sigma^{\prime\prime}}^{\dagger}(\bm{r}_{1},\tau)\psi_{\sigma^{\prime}}^{\dagger}(\bm{r}_{2})}$. 
Defining $\Delta_{\alpha\beta}(\bm{r}) = g\braket{T_{\alpha\beta}(\bm{r})}=g\braket{T_{\alpha\beta}(\bm{r},\tau)}$, we introduce the mean field by $\hat{\Delta}(\bm{r}) =  -\sum_{\alpha\beta}\frac{1}{2}\{\Delta_{\alpha\beta}(\bm{r}),\hat{t}_{\alpha\beta}^{*}(\ii\bar{\bm{\nabla}})\}$. The Gor'kov equation is given  by
\begin{align}
-\check{1}\delta(\bm{r}_{1}-\bm{r}_{2})
= \left[\ii \omega_{n} - \check{H}_{\mathrm{BdG}}(\bm{r}_{1})\right]\check{\tau}_{3}\check{G}(\bm{r}_{1},\bm{r}_{2};\ii\omega_{n}),\\
\check{H}_{\mathrm{BdG}}(\bm{r}) = 
\begin{pmatrix}
h_{0}(\bm{r})\hat{1} + \hat{U}(\bm{r}) & \hat{\Delta}(\bm{r})\\ -  \hat{\Delta}^{*}(\bm{r}) & - h_{0}(\bm{r})\hat{1} - \hat{U}^{\mathrm{T}}(\bm{r})
\end{pmatrix}.
\end{align}
The gap equation is represented using the Green function as
\begin{align}
\Delta_{\alpha\beta}(\bm{R}) 
&
=gT\sum_{n}\lim_{\bm{r}_{2}\to\bm{r}_{1}}\Tr[\hat{t}_{\alpha\beta}(-\ii\bar{\bm{\nabla}}_{1})\hat{F}(\bm{r}_{1},\bm{r}_{2};\ii\omega_{n})]\nonumber \\
&= g T\sum_{n}\int \dfrac{\dd \bm{k}}{(2\pi)^{d}}\Tr[\hat{t}_{\alpha\beta}(\bm{k}/k_{\F})  \hat{F}(\bm{k},\bm{R};\ii\omega_{n})].
\end{align}

\subsection{Eilenberger equation}
In a similar way, we have a right-Gor'kov equation. The quasiclassical transport equation can be obtained by (i) subtracting the right one from the left one, (ii) integrating the equation over the single particle energy $\xi_{\bm{k}} = \frac{k^{2}}{2m} - \mu$, and (iii) retaining the contribution from the pair potentials with Fermi momentum $\bm{k}_{\F}$. We obtain the Eilenberger equation as 
\begin{align}
0 &= \ii \bm{v}_{\F}\cdot\bar{\bm{\nabla}} \check{g}(\bm{k}_{\F},\bm{R};\ii\omega_{n})\nonumber \\ &+ [\ii\omega_{n}\check{\tau}_{3} + \check{u}(\bm{R}) +\check{\sigma}_{\Delta}(\bm{k}_{\F},\bm{R}), \check{g}(\bm{k}_{\F},\bm{R};\ii\omega_{n})],
\label{eq:eilen}
\end{align}
where $\check{u}  =  V_{\mathrm{Z}}\mathrm{diag}(\hat{\sigma}_{3},\hat{\sigma}_{3}^{*})$, and 
\begin{align}
\check{\sigma}_{\Delta} (\bm{k}_{\F},\bm{R})=\begin{pmatrix}
\hat{0} & \hat{\Delta}(\bm{k}_{\F},\bm{R})\\
-\hat{\Delta}^{\dagger}(\bm{k}_{\F},\bm{R}) & \hat{0}
\end{pmatrix},\\
\check{g}(\bm{k}_{\F},\bm{R};\ii\omega_{n}) = \oint_{C_{\mathrm{qc}}} \dfrac{\dd\xi}{\ii \pi}\check{G}(\bm{k},\bm{R};\ii\omega_{n}).
\end{align}
The contour $C_{\mathrm{qc}}$ stands for the mean of two-contour contributions: One is the counterclockwise contour in the half upper $\xi_{\bm{k}}$ plane, and the other is the clockwise contour in the half lower $\xi_{\bm{k}}$ plane\cite{Eilenberger1968,Masaki2019}. 
In the quasiclassical approximation, the gap equation is reduced to
\begin{align}
\hat{\Delta}(\bm{k}_{\F},\bm{R}) = \sum_{\alpha\beta}\Delta_{\alpha\beta}(\bm{R})\hat{t}_{\alpha\beta}^{*}(\bar{\bm{k}}_{\F}),\\
\Delta_{\alpha\beta}(\bm{R}) = g\nu_{\n}\ii \pi T\sum_{n} \braket{\Tr\hat{t}_{\alpha\beta}(\bar{\bm{k}}_{\F})\hat{f}(\bm{k}_{\F},\bm{R};\ii\omega_{n})}_{\F}, \label{eq-gap-eq-qc}
\end{align}
where $\bar{\bm{k}}_{\F} = \bm{k}_{\F} / k_{\F}$ and the quasiclassical anomalous propagator $\hat{f}$ is defined as
\begin{align}
\check{g} (\bm{k}_{\F},\bm{R};\ii\omega_{n}) = \begin{pmatrix}
\hat{g}(\bm{k}_{\F},\bm{R};\ii\omega_{n}) & \hat{f}(\bm{k}_{\F},\bm{R};\ii\omega_{n}) \\ 
-\hat{\bar{f}}(\bm{k}_{\F},\bm{R};\ii\omega_{n}) & \hat{\bar{g}} (\bm{k}_{\F},\bm{R};\ii\omega_{n})
\end{pmatrix}.
\end{align}
The order parameter tensor $A_{\alpha\beta}(\bm{R})$, which is defined as $\hat{\Delta}(\bm{k}_{\F},\bm{R}) = \sum_{\alpha\beta}A_{\alpha\beta}(\bm{R})\hat{\sigma}_{\alpha}\ii\hat{\sigma}_{2}\bar{k}_{\beta}$, can be expressed as follows:
\begin{align}
A_{\alpha\beta}(\bm{R}) = -\dfrac{\Delta_{\alpha\beta}(\bm{R}) + \Delta_{\beta\alpha}(\bm{R})}{2\sqrt{2}}  + \dfrac{\sum_{\gamma}\Delta_{\gamma\gamma}(\bm{R})}{3\sqrt{2}}\delta_{\alpha\beta}.
\end{align}

\subsection{Axisymmetric condition}

\begin{figure*}
\includegraphics[width=55em]{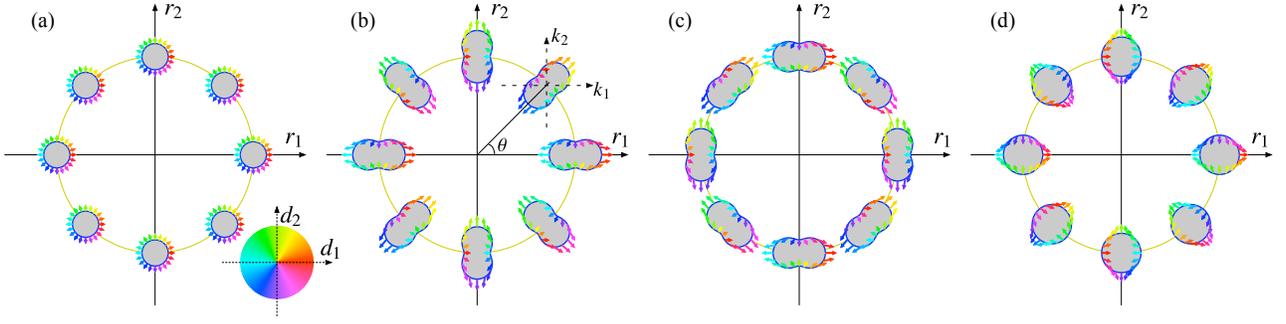}
\caption{Schematic images of boundary conditions using the $d$ vectors without vorticity $\kappa$: $d_{\mu}(\bm{k}_{\F},\theta) = \sum_{j}\sum_{M}\gamma_{M}(\infty)e^{-\ii M \theta}\Gamma_{M,\mu j}\bar{k}_{j}$. The arrows and the blue curves of the objects stand for the directions and amplitude of the $d$ vectors, respectively, at $\bm{k}_{\F}(\bar{k}_{3} = 0)$ and $\theta$, which are indicated, respectively, by the position of the arrows in each object and by the position of the objects (see panel (b)). The arrows are shown with colors, which also show the direction of the $d$ vector for the visibility. Note that the order parameters at the boundary are unitary: $\bm{d}(\bm{k}_{\F},\theta)\times \bm{d}^{*}(\bm{k}_{\F},\theta) = 0$. The largest amplitude points to the direction of the max. EV of $A_{\mu i}$. Panels (a)--(c) represent  boundary conditions at zero magnetic field $V_{\mathrm{Z}}=0$. They are characterized by the direction of the max.~EV of the UN phase. The directions are parallel to (a) the vorticity, and (b) the radial and (c) angular directions in the plane perpendicular to the vorticity. Panel (d) represents the boundary condition in the $D_{4}$-BN phase for field $V_{\mathrm{ZT}} > V_{\mathrm{Zc}}$.}
\label{fig-scheme}
\end{figure*}
This tensor $A$ has a representation using the third component of the angular momentum $M$ as $A(\bm{R}) = \sum_{M=-2}^{2}\gamma_{M}(\bm{R})\Gamma_{M}$, where $\gamma_{M}(\bm{R})$ is a scalar function and $\Gamma_{M}$ is a 3 by 3 tensor. 
The representations of $\Gamma_{M}$ are explicitly shown in Appendix~\ref{sec:RepCooperPair}. The axisymmetric condition, which is given by $(J_{3} -\kappa)A(\bm{R}) = 0$ for the total angular momentum $\kappa$, reads
\begin{align}
A(\bm{R}) = \sum_{M=-2}^{2}\gamma_{M}(\rho)e^{\ii (\kappa - M )\theta}\Gamma_{M}. \label{eq-axi-cond}
\end{align}
The definitions of angular momentum operators are given in Appendix~\ref{sec:AM}. It is instructive to see the tensor in the cylindrical representation, in which $A = R_{\theta}A^{\Cyl}R_{\theta}^{\mathrm{T}}$. Note that $R_{\theta}$ denotes the rotational matrix along the third axis by angle $\theta$.
Since $R_{\theta}\Gamma_{M}R_{\theta}^{\mathrm{T}} = \Gamma_{M} e^{-\ii M \theta}$,  we have 
\begin{align}
A^{\Cyl}(\bm{R}) &= \sum_{M=-2}^{2}\gamma_{M}(\rho)e^{\ii\kappa\theta}\Gamma_{M}^{\Cyl},
\end{align}
where $\Gamma_{M}^{\Cyl}$ has the same representation as $\Gamma_{M}$ but for basis $\bm{k}_{\F} = k_{\rho}\bm{e}_{\rho} + k_{\theta}\bm{e}_{\theta} + k_{3}\bm{e}_{3}$, and $\hat{\bm{\sigma}} = \hat{\sigma}_{\rho}\bm{e}_{\rho} + \hat{\sigma}_{\theta}\bm{e}_{\theta} + \hat{\sigma}_{3}\bm{e}_{3}$ and
\begin{align}
k_{\rho} =  \cos\theta k_{1}+\sin\theta k_{2},\\
k_{\theta} = -\sin\theta k_{1}+\cos\theta k_{2},\\
\hat{\sigma}_{\rho} =  \cos\theta\hat{\sigma}_{1}+\sin\theta\hat{\sigma}_{2},\\
\hat{\sigma}_{\theta} = -\sin\theta\hat{\sigma}_{1}+\cos\theta\hat{\sigma}_{2}.
\end{align}
The rotation of the order parameter is also discussed in Appendix~\ref{sec:Rotation}. 
There are several choices of $\gamma_{M}(\rho\to \infty)$, and 
the schematic pictures of representative cases are shown in Fig.~\ref{fig-scheme},
using the $d$ vectors without vorticity $\kappa$: $d_{\mu}(\bm{k}_{\F},\theta) = \sum_{j}\sum_{M}\gamma_{M}(\infty)e^{-\ii M \theta}\Gamma_{M,\mu j}\bar{k}_{j}$.
 Panels (a)--(c) are for the UN phase, while panel (d) is for the $D_{4}$-BN phase

\subsection{Free energy}
We calculate the free energy on the basis of the Luttinger--Ward formalism.
By solving the Eilenberger equation combined with the gap equation, we have determined the self energy $\check{\sigma}$ self-consistently.
Let us define an auxiliary Green function in the Gor'kov equation as
\begin{align}
\check{G}_{\lambda}^{-1}(\bm{r}_{1},\bm{r}_{2};\ii\omega_{n})  = \check{G}_{0}^{-1}(\bm{r}_{1},\bm{r}_{2};\ii\omega_{n}) - \lambda\check{\Sigma}(\bm{r}_{1},\bm{r}_{2};\ii\omega_{n}).
\end{align}
Note that $\check{G}_{\lambda=0} = \check{G}_{0}$ and $\check{G}_{\lambda=1} = \check{G}$. 
Following Ref.~[\onlinecite{Vorontsov2003}], we obtain the difference of the thermodynamic potentials between the superfluid and normal states as
\begin{align}
\mathcal{J}_{\mathrm{sn}}  = \dfrac{\nu_{\n}}{2}\int_{0}^{1} \dd \lambda \mathrm{Sp}\left[\check{\sigma}_{\Delta}\left(\check{g}_{\lambda} - \dfrac{1}{2}\check{g}\right)\right],
\end{align}
where  $\mathrm{Sp} [\cdots]= \ii \pi T\sum_{n} \int \dd \bm{R}\int \frac{\dd \hat{\bm{k}}}{4\pi}\Tr[\cdots]$ and $\check{g}_{\lambda}$ is the solution to the equation
\begin{align}
[\ii\omega_{n} \check{\tau}_{3} + \check{u} + \lambda \check{\sigma}_{\Delta},\check{g}_{\lambda}] = 0.
\end{align}

When the system is axisymmetric, we obtain the free-energy density as a function of $\rho$ with $\Omega = \pi R^{2} \Omega_{3}$
\begin{align}
\dfrac{ \mathcal{J}_{\mathrm{sn}}}{\Omega} 
&= \dfrac{\nu_{\n}T_{\mathrm{c}}^{2}}{(R/\xi_{0})^{2}}
\int \dd \bar{\rho} \bar{\rho}\bar{\mathfrak{J}}_{\mathrm{sn}}(\rho)\equiv\dfrac{\nu_{\n}T_{\mathrm{c}}^{2}}{(R/\xi_{0})^{2}}\bar{\mathcal{J}}_{\mathrm{sn}},\\
\bar{\mathfrak{J}}_{\mathrm{sn}}(\rho) &\equiv \int_{0}^{1}\dd \lambda\dfrac{\ii \pi T}{T_{\mathrm{c}}^{2}}\sum_{n}
\Tr\left\langle
\hat{\bar{\Delta}} \left(\hat{f}_{\lambda} -\frac{\hat{f}}{2}\right)
- \hat{\Delta}\left(\hat{\bar{f}}_{\lambda}-\frac{\hat{\bar{f}}}{2}\right)
\right\rangle_{\F},
\end{align}
where $\Tr$ in the second line stands for the trace of the 2 by 2 matrix in the spin space. We show the dimensionless total free energy $\bar{\mathcal{J}}_{\mathrm{sn}}$ and free-energy density $\bar{\mathfrak{J}}_{\mathrm{sn}}(\rho)$ in Sect.~\ref{sec:results}.

\subsection{Bogoliubov--de Gennes equation}
The BdG equation is derived from the equation $\ii \partial_{t}\psi_{\sigma}^{(\dagger)} = [\psi_{\sigma}^{(\dagger)},H]$ using a mean field approximation and $\vec{\Psi}(\bm{r},t) = \sum_{\nu}e^{-\ii E_{\nu}t}\vec{u}_{\nu}(\bm{r})\alpha_{\nu}$.
\begin{align}
E_{\nu} \vec{u}_{\nu}(\bm{r}) &=\check{\mathcal{H}}_{\BdG} (\bm{r})\vec{u}_{\nu}(\bm{r}),
\label{eq:bdg}
\\
\vec{u}_{\nu}(\bm{r}) &= {}^{t}[u_{\uparrow,\nu}(\bm{r}),u_{\downarrow,\nu}(\bm{r}),v_{\uparrow,\nu}(\bm{r}),v_{\downarrow,\nu}(\bm{r})].
\end{align}
Here we note that 
\begin{align}
\hat{\Delta}(\bm{r}) &= \dfrac{1}{2}\sum_{\alpha\beta}\{\Delta_{\alpha\beta}(\bm{r}),\hat{t}_{\alpha\beta}(-\ii\bar{\bm{\nabla}})\} \nonumber \\
&= \dfrac{1}{2}\sum_{\alpha\beta}\{A_{\alpha\beta}(\bm{r}),(-\ii\bar{\bm{\nabla}})_{\beta}\}\hat{\sigma}_{\alpha}\ii\hat{\sigma}_{2}\nonumber \\
&= \dfrac{1}{2}\sum_{\alpha\beta}\sum_{M}\{\gamma_{M}(\bm{r}),(-\ii\bar{\bm{\nabla}})_{\beta}\}\Gamma_{M,\alpha\beta}\hat{\sigma}_{\alpha}\ii\hat{\sigma}_{2}.\nonumber 
\end{align}
We use the cylindrical coordinate $(\rho,\theta,r_{3})$ and note the following: 
\begin{align}
\dfrac{\partial}{\partial r_{1}} \pm \ii \dfrac{\partial}{\partial r_{2}} = e^{\pm \ii \theta}\left( \dfrac{\partial}{\partial \rho} \pm\ii \dfrac{1}{\rho}\dfrac{\partial}{\partial \theta}\right).
\end{align}
The eigenvector can be given by $\vec{u}_{\nu}(\rho,\theta, r_{3}) =\frac{e^{\ii k_{3} r_{3}}}{\sqrt{\Omega_{3}}}\check{U}_{\ell,\kappa}(\theta)\vec{u}_{n,\ell,k_{3}}(\rho)$ with a 4 by 4 diagonal matrix $\check{U}_{\ell,\kappa}(\theta) =\frac{1}{\sqrt{2\pi}} \mathrm{diag}(e^{\ii \ell \theta},e^{\ii (\ell+1)\theta}, e^{\ii(\ell+1-\kappa)\theta},e^{\ii(\ell-\kappa)\theta})$ and the radial part to be determined:
\begin{align}
\vec{u}_{\nu= (n,\ell,k_{3})}(\rho) =\begin{pmatrix}
 u_{\uparrow n,\ell,k_{3}}(\rho)\\
 u_{\downarrow n,\ell+1,k_{3}}(\rho)\\
 v_{\uparrow n,\ell+1-\kappa,k_{3}}(\rho)\\
 v_{\downarrow n,\ell-\kappa,k_{3}}(\rho)
\end{pmatrix},
\end{align}
because of the equations $[\check{\mathcal{H}}_{\BdG}, J_{3}\check{1}-\frac{\kappa}{2}\check{\tau}_{3}]=0$ and $[\check{\mathcal{H}}_{\BdG}, (-\ii \partial_{3})\check{1}]=0$.  We may use the Bessel functions and their zeros for expansion of the radial part ($w = u$ or $\varv$):
\begin{align}
w_{\sigma \nu}(\rho)  = \sum_{k} \phi_{\ell,k}(\rho) w_{\sigma \nu,k},\\
\phi_{\ell,k}(\rho) = \dfrac{\pm\sqrt{2}J_{\ell}(\beta_{\ell,k}\rho/R_{0})}{R_{0}J_{\ell\pm1}(\beta_{\ell,k})}.
\end{align}
Here $\beta_{\ell,k}$ denotes the $k$ th zero of the Bessel function $J_{\ell}$. The orthonormalization condition is given by
\begin{align}
\int_{0}^{R_{0}} \dd \rho \rho \phi_{\ell,k}(\rho) \phi_{\ell,k^{\prime}}(\rho) = \delta_{k,k^{\prime}}.
\end{align}
We obtain the Hamiltonian matrix to be diagonalized as 
$[\check{\mathcal{H}}_{\BdG}(\ell,k_{3})]_{k,k^{\prime}}= \int_{0}^{R_{0}} \dd \rho \rho \check{\Phi}_{\ell,k}(\rho) \check{\mathcal{H}}_{\BdG}(\rho,\ell,k_{3})\check{\Phi}_{\ell,k^{\prime}}(\rho)$
with $\check{\Phi}_{\ell,k}(\rho) = \mathrm{diag}(\phi_{\ell,k},\phi_{\ell+1,k},\phi_{\ell+1-\kappa,k},\phi_{\ell-\kappa,k})(\rho)$, where the eigen equation takes the following form:
\begin{align}
\sum_{k^{\prime}}[\check{\mathcal{H}}_{\BdG}(\ell,k_{3})]_{k,k^{\prime}} \vec{u}_{\nu,k^{\prime}} =  E_{\nu}\vec{u}_{\nu,k}.
\end{align}
We remark that there is a relation between the positive and negative eigenvalues. When we fix $\kappa = 1$, the relation reads the one between the state with $\nu = (n,\ell,k_{3})$ and the state with $\bar{\nu} = (\bar{n},-\ell,-k_{3})$ for some $\bar{n}$:
\begin{align}
E_{\bar{\nu}} &= -E_{\nu},~\vec{u}_{\bar{\nu}}(\bm{r}) = [\check{\tau}_{1}\vec{u}_{\nu}(\bm{r})]^{*}.
\end{align}
The Bogoliubov transformation is given by 
\begin{align}
\psi_{\sigma}(\bm{r})  
= \sum_{\nu: E_{\nu} > 0}
\left(u_{\sigma,\nu}(\bm{r})\alpha_{\nu} + v_{\nu}^{*}(\bm{r}) \alpha_{\nu}^{\dagger}\right).
\end{align}
We calculate the expectation value of the third component of local spin $S_{3}(\bm{r}) = \frac{1}{2}(\psi_{\uparrow}^{\dagger}(\bm{r})\psi_{\uparrow}(\bm{r})-\psi_{\downarrow}^{\dagger}(\bm{r})\psi_{\downarrow}(\bm{r}))$ as
\begin{align}
\braket{S_{3}(\bm{r})} &= \dfrac{1}{2}\sum_{\nu: E_{\nu}>0}\left[
(|u_{\uparrow\nu}(\bm{r})|^{2}-|u_{\downarrow\nu}(\bm{r})|^{2})f(E_{\nu})\right.\nonumber \\&\hspace{2em}\left.
+(|v_{\uparrow\nu}(\bm{r})|^{2}-|v_{\downarrow\nu}(\bm{r})|^{2})(1-f(E_{\nu}))\right]
, \label{eq-spin}
\end{align}
where $f(E_{\nu}) = \braket{\alpha_{\nu}^{\dagger}\alpha_{\nu}}=1/(e^{E_{\nu}/T}+1)$ is the Fermi distribution function.

\section{Numerical results}\label{sec:results}

In this section, by self-consistently solving the Eilenberger Eq.~\eqref{eq:eilen} and gap Eq.~\eqref{eq-gap-eq-qc} with the boundary conditions as shown in Fig.~\ref{fig-scheme}, we show the core structure of stable $o$ and $v$ vortices in \tPt superfluids. Using the BdG Eq.~\eqref{eq:bdg} with the gap function obtained from the quasiclassical theory, we discuss excitation spectra and magnetizations around the vortices. We set $T=0.4T_{\mathrm{c}}$ and $\omega_{\mathrm{c}} = 10 T_{\mathrm{c}}$ for all the calculations shown here.
The units of energy and length are, respectively, $T_{\mathrm{c}}$ and $\xi_{0} = \varv_{\F}/(2\pi T_{\mathrm{c}})$. 
\begin{figure*}[t]
\includegraphics[width = \hsize]{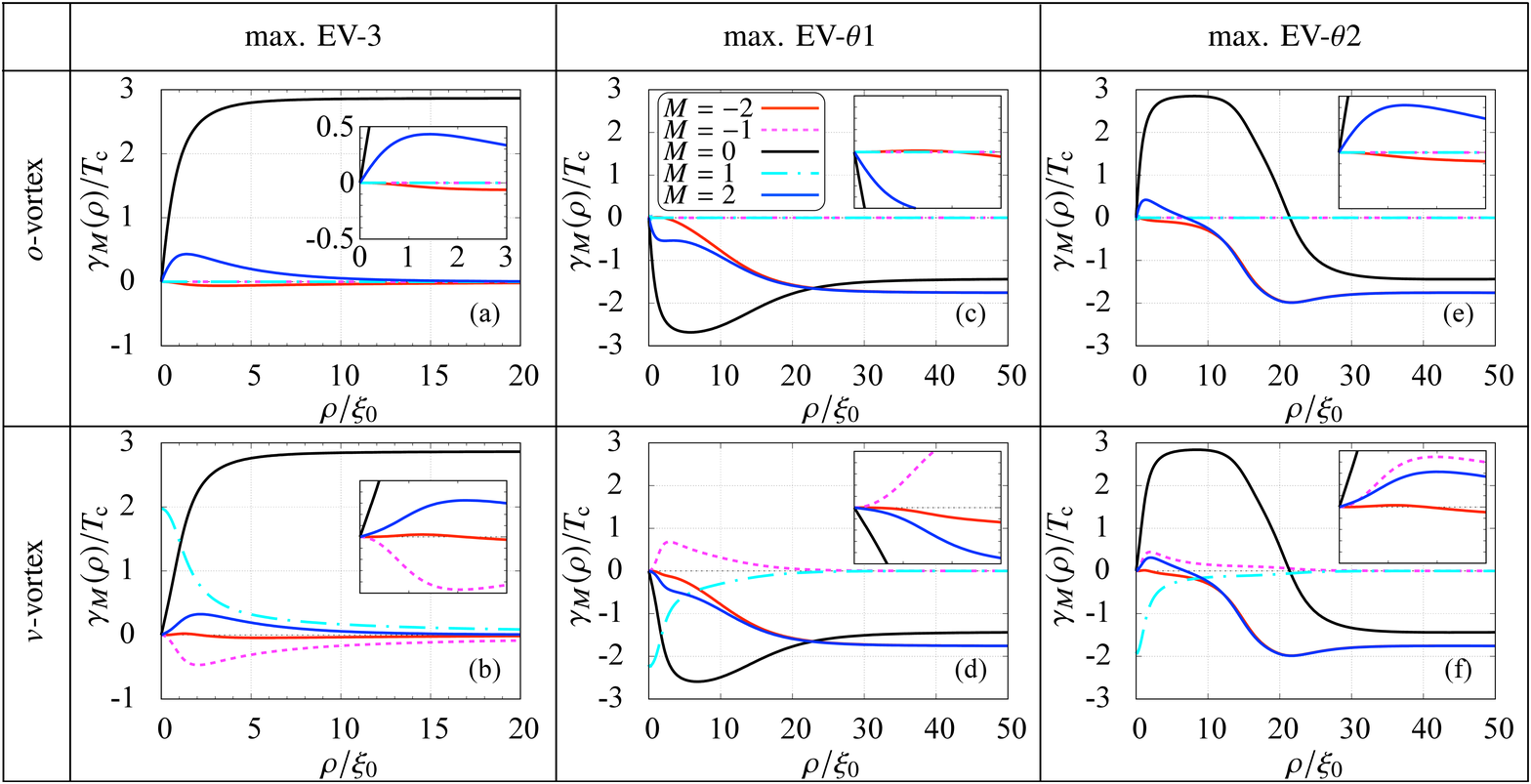}
\caption{
Order parameter profiles of (a), (c), (e) $o$ vortices and (b), (d), (f) $v$ vortices in UN states. At the boundary, the max. EV of the UN state points to the third direction as shown in Fig.~\ref{fig-scheme}(a) for panels (a) and (b), while it points to the angular direction as shown in Fig.~\ref{fig-scheme}(c) for panels (c)--(f)[(c, d) $\theta1$ vortex, and (e, f) $\theta2$ vortex, see the text].  The relation between the color and the component $\gamma_{M}(\rho)$ is available in panel (c). 
The inset in each panel shows the profile around the vortex core $\rho \le 3\xi_{0}$.
The components in $o$ vortices are slightly different from those in $v$ vortices owing to the induced components $\gamma_{M=\pm1}$. 
}
\label{fig-UN-gap}
\end{figure*}

Since we show several kinds of single vortices, we summarize our classification rule here.
Their differences appear owing to the symmetry of order-parameter components around the core and the boundary conditions. Regarding the symmetry around the core, we construct $o$ and $v$ vortices; They are distinguished whether a $P_{3}$ symmetry exists or not. This classification is based on the context of the superfluid ${}^{3}$He-B phase\cite{Salomaa1987}, and  details are discussed in the next subsection. As for the boundary conditions, we label $3$, $\rho$, $\theta$ for vortices in the UN or $D_{2}$-BN phase, which depend on the boundary conditions, while use no label for the $D_{4}$-BN phase. Let $V_{\mathrm{Zc}}$ be a transition magnetic field between the $D_{2}$-BN and $D_{4}$-BN states. For magnetic field $V_{\mathrm{Z}} < V_{\mathrm{Zc}}$, where either UN or $D_{2}$-BN state realizes, the name is determined by the direction of the maximum eigenvalue (max.~EV) of the order parameter tensor. We consider three representative directions: the third, radial and angular directions.  At zero magnetic field, the schematic images of the boundary conditions for these three cases are shown in Figs.~\ref{fig-scheme}(a)--\ref{fig-scheme}(c), which stand for $3$, $\rho$, and $\theta$ vortices, respectively. In the presence of magnetic field parallel to the third axis, the boundary conditions are obtained by transforming Figs.~\ref{fig-scheme}(b) and \ref{fig-scheme}(c) continuously, though we do not explicitly show their schematic images. Note that when the magnetic field is parallel to the direction of the max. EV, such an order parameter is unstable and changes its direction of the max. EV. Therefore we do not study magnetic field effects on the $3$ vortex. For $V_{\mathrm{Z}}\ge V_{\mathrm{Zc}}$, the $D_{4}$-BN state is realized and the boundary conditions of $\rho$ and $\theta$ become the same. We name vortices in that region ``$D_{4}$-BN vortex''.
When we have several configurations with the same boundary condition, we just add further labels 1 and 2 to the above rules to distinguish them, e.g., $\theta1$ and $\theta2$ and so on.

\subsection{Zero magnetic field}
First we consider the zero-field case, where the equilibrium state is the UN state. The state can be characterized by the direction of the max.~EV of the order parameter matrix $A$. 
A simple form of single vortex states is given when the max.~EV points to the third direction, i.e., $A = A_{0}\mathrm{diag}(-1/2,-1/2,1)$ [Fig.~\ref{fig-scheme}(a)], which we call the $3$ vortex. The boundary condition is imposed at $\rho = R_{\mathrm{c}}$ as 
$A(R_{\mathrm{c}},\theta) = \gamma_{0}e^{\ii \kappa \theta}\Gamma_{0}$ with the vorticity $\kappa$, where $\gamma_{0}$ is the value of the order parameter in the uniform state. We fix $\kappa=1$ and first show the possible axisymmetric vortex solutions and compare the results with those for ${}^{3}$He-B. In the earlier work using GL theory up to the sixth order term\cite{Masuda:2015jka}, the axisymmetric form of $A(\bm{R}) = \gamma_{0}(\rho)e^{\ii \kappa \theta}\Gamma_{0}$ is only considered. We investigate other solutions in the form of Eq.~\eqref{eq-axi-cond} in an analogy to the case of ${}^{3}$He-B.

\begin{figure*}[t]
\includegraphics[width = \hsize]{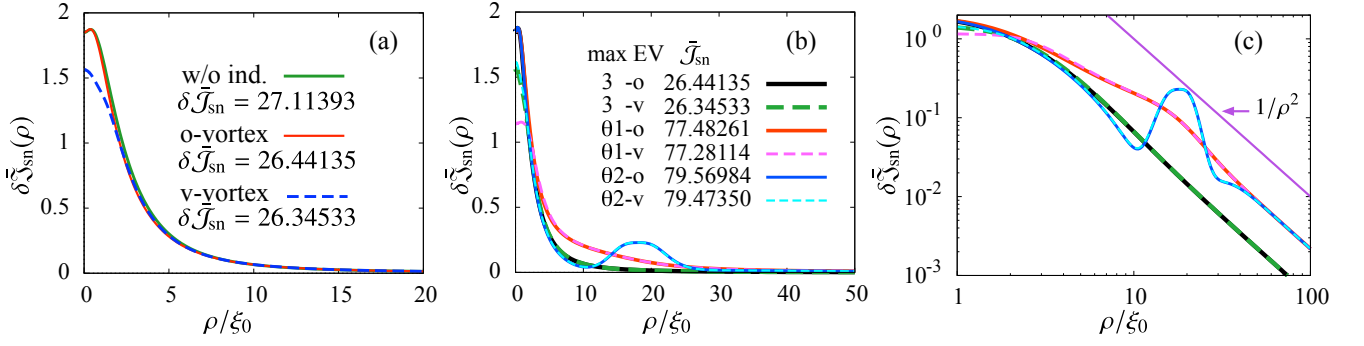}
\caption{
(a) Free-energy densities for vortices with max. EV parallel to the third direction. The integrated values from 0 to $R_{\mathrm{c}} \simeq 100$ is given in the legend.
The curves labeled as ``$o$-vortex'', and ``$v$-vortex'' are obtained on the basis of the order parameter profiles shown in Figs.~\ref{fig-UN-gap}(a) and \ref{fig-UN-gap}(b), respectively. The curve labeled as ``w/o ind.'' is obtained for $\gamma_{M=0}(\rho)$ shown in Fig.~\ref{fig-UN-gap}(a) with $\gamma_{M\neq0} = 0$. 
(b) Free-energy densities for several order parameter configurations. The labels from the top to the bottom, showing the corresponding order parameter profile,  are put in the same order as the panels in Figs.~\ref{fig-UN-gap}(a) -- \ref{fig-UN-gap}(f).  Their integrated values are also shown in the legend. (c) Free-energy densities in the logarithmic scale with $1/\rho^{2}$ behavior for an eye-guide. The color plots are the same as those in panel (b). 
}
\label{fig-UN-free-energy}
\end{figure*}
Figures~\ref{fig-UN-gap}(a) and \ref{fig-UN-gap}(b) show the spatial profiles of the order parameter. 
The axisymmetric form of $A(\bm{R}) = \gamma_{0}(\rho)e^{\ii \kappa \theta}\Gamma_{0}$ in Ref.~[\onlinecite{Masuda:2015jka}] corresponds to $\gamma_{M=0}(\rho)$ in the panel (a) with $\gamma_{M=\pm 2} = 0$.
We remark that it does not satisfy the gap Eq.~\eqref{eq-gap-eq-qc} since the right-hand side in Eq.~\eqref{eq-gap-eq-qc}  has nonzero components with $M=\pm2$.
Instead, we obtain the profile shown in panel (a) with nonzero induced components $\gamma_{M=\pm2}(\rho)$, which is the so-called ``$o$ vortex'' in the context of $^{3}$He B-phase. The $o$ vortex is the most symmetric vortex in terms of the three discrete symmetries: They are called $P_{1}$, $P_{2}$, and $P_{3}$ symmetries with $P_{1}P_{2}P_{3} = 1$. The physical meaning of these symmetries are, respectively, inversion, magnetic mirror, and magnetic $\pi$ rotation symmetries and the details are referred to as Refs.~[\onlinecite{Salomaa1985,Salomaa1987,Tsutsumi2015,mizushimaJPSJ16}]. The presence of these symmetries is equivalent for \tPt superfluids to the case in which the components of $M=\pm2$ and $0$ are real, and those of $M=\pm1$ are zero.  Perturbations with nonzero $\gamma_{\pm1}$ change it to ``$v$ vortex'' shown in panel (b). The $v$ vortex is characterized by the presence of $P_{2}$ symmetry and broken $P_{1}$ ($P_{3}$) symmetry. It is represented by five real components $\gamma_{M=-2,\cdots,2}$.  Since the $v$ vortex has the unwinding component, $M=1$, the vortex core is filled with $\gamma_{1}$. The power law of $\gamma_{\pm1}$ in the asymptotic region is $1/\rho$, while $\gamma_{0,\pm2}$ approaches to the bulk values with $1/\rho^{2}$. This asymptotic behavior is the same as that of the $v$ vortex in the $^{3}$He B-phase. The difference may appear owing to the restriction of the total angular momentum to the $J=2$ sector. The $v$ vortex in the $^{3}$He B-phase is considered to have the A-phase core since the A-phase component at $\rho = 0$ is more dominant than the $\beta$-phase component. Only these two components are allowed to be nonzero at the origin, and they have opposite signs. On the other hand, in the case of \tPt superfluids, these two components always have the same magnitude with the same sign because the order parameter tensor is symmetric. We also calculate free energy densities of these vortices.  Figure~\ref{fig-UN-free-energy}(a) shows the free energy densities of these vortices measured from the uniform solution, i.e., $\delta \mathfrak{J}_{\mathrm{sn}}(\rho) \to 0 (\rho\to \infty)$ with $1/\rho^{2}$. The integrated values shown in its legend imply that the $v$ vortex is the most stable. 
It seems to gain the condensation energy at the vortex center by filling its core with $\gamma_{1}$. The $o$ vortex has less free energy owing to the condensation energy of $\gamma_{\pm2}$ than the vortex constructed by $\gamma_{0}$ without any induced components, i.e., $\gamma_{M\neq 0} = 0$, which is labeled ``w/o ind.'' in Fig.~\ref{fig-UN-free-energy}(a).

\begin{figure*}[t]
\includegraphics[width = \hsize]{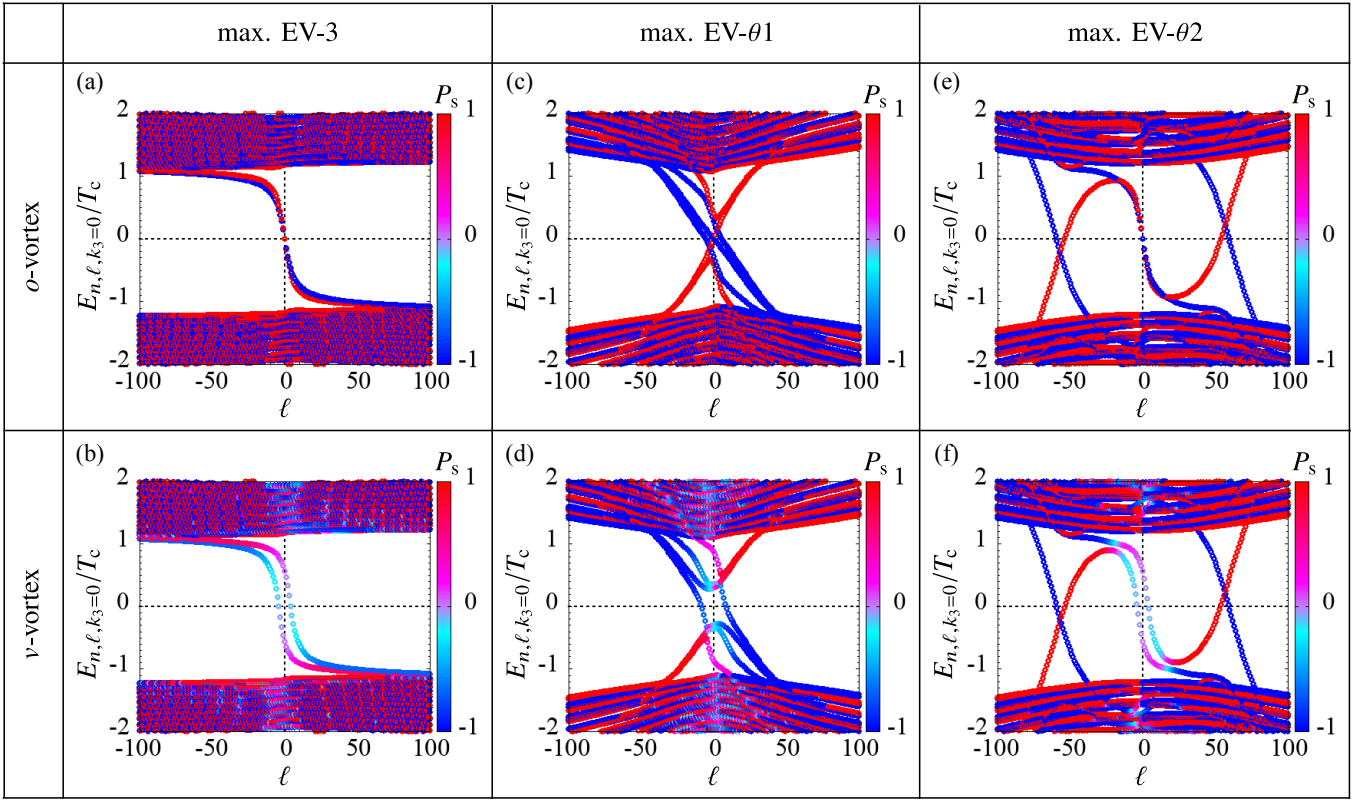}
\caption{Eigen spectra obtained by solving BdG equations.  Each panel shows the spectra for the corresponding vortex with the profile in Fig.~\ref{fig-UN-gap}.
We input the pair potentials obtained as self-consistent solutions to the quasiclassical equations into the BdG equation with a quasiclassical parameter  $k_{\F}\xi_{0} = 4$. For interests in Majorana fermions, we show the eigenspectra for $k_3 = 0$.
A horizontal (vertical) axis is the angular momentum (eigenenergy) of the quasiparticle, while the color plot stands for 
the spin polarization rate as defined in the main text. 
}
\label{fig-BdG-eigen}
\end{figure*}
We show other possible boundary conditions of axisymmetric vortices in the absence of the magnetic field. They are more relevant for finite magnetic field because 
the above vortex states are unstable against the magnetic field parallel to the third axis. We consider two representative directions of the max.~EVs which are the radial and angular directions [Figs.~\ref{fig-scheme}(b) and \ref{fig-scheme}(c), respectively]. At zero field, we construct solutions for the angular direction only and we will discuss the case of the radial direction later in the presence of the magnetic field. As far as we investigated, we have found two kinds of solutions $\theta1$ (Figs.~\ref{fig-UN-gap}(c) and \ref{fig-UN-gap}(d)) and $\theta2$ (Figs.~\ref{fig-UN-gap}(e) and \ref{fig-UN-gap}(f)), each of which has vortices with a core ($o$ vortex) and without a core ($v$ vortex).  We label panels (c)--(f) ``$\theta1$-o'', ``$\theta2$-o'', ``$\theta1$-v'', and ``$\theta2$-v'', respectively. In all the cases, there are structures similar to a domain wall ring in $10\lesssim \rho/\xi_{0} \lesssim 30$, and structures in the domain including the vortex center are well-described by those in Figs.~\ref{fig-UN-gap}(a) and \ref{fig-UN-gap}(b). In the middle and right panels in Fig.~\ref{fig-UN-gap}, the order parameters in the domains including their vortex cores have opposite signs. The order parameters around the domain wall rings are also different. In the middle panels in Fig.~\ref{fig-UN-gap}, $\gamma_{0}(\rho)$ does not cross the zero values. We discuss which vortices are the most stable on the basis of the free energy. The free-energy density for each profile is shown in Fig.~\ref{fig-UN-free-energy}(b). We see that the free energies of the $\theta1$ vortices (Figs.~\ref{fig-UN-gap}(c) and \ref{fig-UN-gap}(d)) are lower than those of the $\theta2$ vortices (Figs.~\ref{fig-UN-gap}(e) and \ref{fig-UN-gap}(f)). For the $\theta1$ vortices, the core region and the bulk region are gradually connected, accompanied by loss of the free energy with a gentle tail, as indicated by the red-solid, and pink-dashed curves in Fig.~\ref{fig-UN-free-energy}. The $\theta2$-$o$ and $\theta2$-$v$ vortices have bump structures around the connecting regions in their free energy densities, as indicated by the blue-solid, and light-blue-dashed curves. The difference in these structures determine the free-energy difference between the $\theta1$  and $\theta2$ vortices.
The difference between $o$ and $v$ vortices can be seen in the vortex core, where a $v$ vortex has the condensation energy. In Fig.~\ref{fig-UN-free-energy}(c), the power law in the asymptotic region is $1/\rho^{2}$ and its coefficient depends on the boundary condition, i.e., it does not depend on induced components or the difference in profiles around the cores. 
This hydrodynamic part proportional to $1/\rho^{2}$ gives a logarithmic divergence $\log R_{\mathrm{c}}$ in the free energy $\delta \mathcal{J}_{\mathrm{sn}}$. Therefore, the 3 vortex has much lower energy than any of $\theta$ vortices in the absence of magnetic fields, although it may be unstable in the presence of a magnetic field.

Among all possible vortices, the $o$-vortex states have topologically-protected zero-energy modes bound to vortices regardless of the boundary condition at $\rho \rightarrow \infty$. Following Refs.~[\onlinecite{Tsutsumi2015,mizushimaJPSJ16,shiozakiPRB14}], we start with the semiclassical approximation where the Hamiltonian varies slowly in the real-space coordinate. The spatial modulation due to a vortex line is considered as adiabatic changes in the Hamiltonian as a function of the real-space coordinate surrounding the defect with an angle $\theta$. Then, the Hamiltonian is obtained in the base space, $(\bm{k},\theta)$, as 
\begin{align}
\check{\mathcal{H}}({\bm k},\theta) 
= \begin{pmatrix}
\hat{\varepsilon}({\bm k}) & \hat{\Delta}({\bm k},\theta) \\
\hat{\Delta}^{\dag}({\bm k},\theta) & -\hat{\varepsilon}^{\rm T}(-{\bm k})
\end{pmatrix},
\end{align}
where $\varepsilon({\bm k})$ is composed of the kinetic energy $h_0\hat{1}$ and the Zeeman energy $\hat{U}$. $\hat{\Delta}({\bm k},\theta)$ is the asymptotic form of the vortex order parameter at $\rho \rightarrow \infty$ that satisfies the boundary condition in Fig.~\ref{fig-scheme}. The $o$ vortex preserves the $P_3$ symmetry, that is, the magnetic $\pi$ rotation symmetry. Then, one can construct the chiral operator $\check{\Gamma}$ as a combination of the particle-hole exchange operator and $P_3$ operator, and the BdG Hamiltonian $\check{\mathcal{H}}({\bm k},\theta)$ preserves the chiral symmetry, $\{\check{\Gamma},\check{\mathcal{H}}({\bm k},\theta) \}=0$ for $k_2=k_3=0$. As long as the chiral symmetry is preserved, one can define the one-dimensional winding number for $\theta$ as 
\begin{align}
w_{\rm 1d}(\theta) = -\frac{1}{4\pi \ii} \int \dd k_z {\rm tr}\left[
\check{\Gamma}\check{\mathcal{H}}({\bm k},\theta)\partial _{k_1} \check{\mathcal{H}}({\bm k},\theta)
\right]_{k_2=k_3=0},
\end{align}
where $w_{\rm 1d}(\theta=0) = 2$ and $w_{\rm 1d}(\theta=\pi) = -2$ for the $o$-vortex state regardless of the boundary condition. Then, the topological invariant is defined as the difference of $w_{\rm 1d}(\theta)$ 
\begin{align}
w_{\rm 1d} = \frac{w_{\rm 1d}(0)-w_{\rm 1d}(\pi)}{2} = 2,
\label{eq:w1d}
\end{align} 
which ensures the presence of the two zero energy states at $k_3=0$. Hence, a pair of zero energy states is guaranteed by the $P_3$ symmetry in the $o$-vortex state.

\begin{figure*}
\includegraphics[width = \hsize]{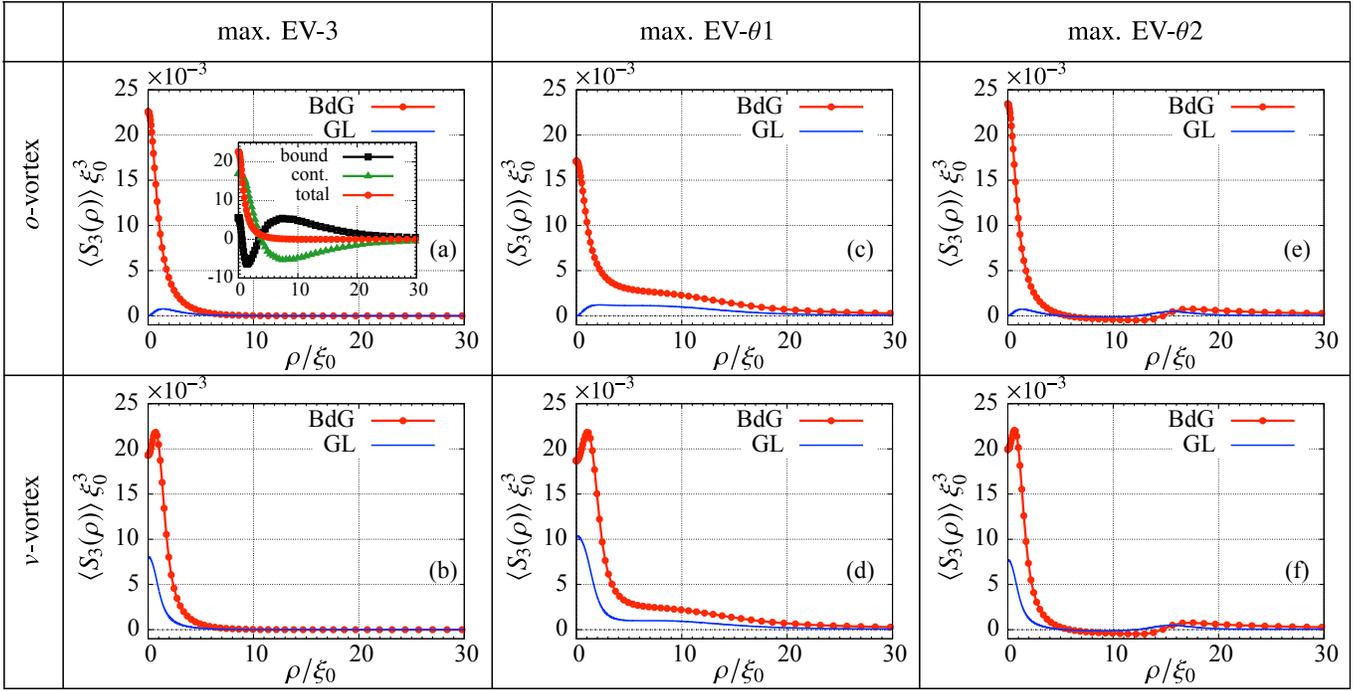}
\caption{Profiles of spin densities.  Each panel shows the spin densities for the corresponding vortex with the profile shown in Fig.~\ref{fig-UN-gap}. The red curves with solid circles are obtained on the basis of Eq.~\eqref{eq-spin}, while the blue curves are calculated using order parameter profiles. The inset in panel (a) shows the spatial profiles of the bound state contribution (black-open squares) and the continuum states contribution (green-solid squares). They compose the total spin profile equivalent to the red circles in the main plot.
}
\label{fig-BdG-spin}
\end{figure*}
Solving the BdG equation \eqref{eq:bdg}, we here investigate the excitation spectra and core magnetizations for the $o$ and $v$ vortices in the cases of max.~EV 3, $\theta1$, and $\theta2$. The effects of magnetic field are discussed in the next subsection. Here we set $k_{\F}\xi_{0} = 4$ and $R_{\mathrm{0}}/\xi_{0} = 80$. Figure~\ref{fig-BdG-eigen} shows the energy spectra of quasiparticles with $k_{3} = 0$. The local density of states (LDOS), which is comparable with the eigen spectra, is shown in Appendix~\ref{sec:ldos-ev} for $\theta2$ vortices. We do not show the helical edge modes which appear in all the cases. Another common feature is that there are two spin degenerated Majorana zero modes with $\ell =0$ in $o$-vortex cores, while they mix up and split in $v$-vortex cores, as is known in $^{3}$He-B, and the anomalous branches cross at $\ell = \pm \ell_{\mathrm{c}}$. The presence of the zero-energy modes is consistent with topological consideration shown in Eq.~\eqref{eq:w1d}.
In addition, the cases of $o$ and $v$ vortices with max.~EV along the third direction (in Figs.~\ref{fig-BdG-eigen}(a) and \ref{fig-BdG-eigen}(b), respectively) are almost the same as those of $^{3}$He-B. Here we show the spin polarization rate using the color bar, which is defined by
\begin{align}
P_{\mathrm{s}} = \int_{0}^{R_{\mathrm{0}}} \dd \rho \rho
\sum_{\sigma}\sigma(|u_{\nu,\sigma}(\rho)|^{2} + |\varv_{\nu,\sigma}(\rho)|^{2}).
\end{align}
The Hamiltonian is block diagonalized by spin sectors $P_{\mathrm{s}} = \pm1$ for the $o$ vortex at $k_{3}=0$. 
In the ${}^{3}$He-B phase, the spin-splitting of the vortex core bound states has been 
found in Refs.~\onlinecite{Silaev2009,Khaymovich2010}.
On the other hand, for the $v$ vortex
$P_{\mathrm{s}}$ takes neither $-1$ nor $1$ around small $\ell$ because the mixing of the spin sectors is caused by $\gamma_{1}$ and $\gamma_{-1}$ which are induced in the $v$-vortex core.

Then we see the case of max. EV-$\theta1$, shown in the middle panels. In Fig.~\ref{fig-BdG-eigen}(c), there are three chiral anomalous branches with spin down crossing at $\ell=0$ and $\ell \simeq\pm 6$. The only one branch appears for the spin up component with the opposite angular velocity, which is the slope at $\ell = 0$. The induced components with $\gamma_{M = \pm1}$ gap out  a pair of chiral branches with spin-up and spin-down crossing at $\ell = 0$, as seen in Fig.~\ref{fig-BdG-eigen}(d). The other two chiral anomalous branches with mainly spin down component are present, but they do not possess topological zero modes.

In the case of max. EV-$\theta2$, excitation spectra with small angular momentum, $|\ell| \lesssim 15$, are similar to those of max. EV-$3$. The sign of the angular velocities of two spin sectors near $\ell=0$ are the same. Topological structures of the anomalous branches are the same as those of max. EV-$\theta1$; as $\ell$ decreases (increases), three branches (a single branch) carry(carries) spin-down (spin-up) quasiparticles from negative energy to positive energy (see Figs.~\ref{fig-BdG-eigen}(c) and \ref{fig-BdG-eigen}(e)). In the presence of $\gamma_{M=\pm1}$ components, two of the anomalous branches turn to two branches which cross zero of energy an even number of times. This topological structure is the same as that in Fig.~\ref{fig-BdG-eigen}(d). 

In summary of the excitation spectra, the adiabatic pumping under a virtual process with increasing an angular momentum, 
two quasiholes are carried into the negative energy region as a net change in any case. In terms of spins, two spin down holes are created 
for max.~EV-$\theta1$ and $\theta2$, while nothing changes for max.~EV-$3$.

Next we see the polarization of spins around the vortex core. We calculate it using Eq.~\eqref{eq-spin} in addition to the calculation using the formula given by
\begin{align}
\braket{S_{3}(\bm{r})} 
&= \dfrac{1}{3}\nu_{\n}^{\prime}(\epsilon_{\F})\ln\left(1.13\dfrac{\Omega_{\mathrm{BCS}}}{T_{\mathrm{c}}}\right)\sum_{\nu=\pm,0}(|A_{+\nu}^{\mathrm{AM}}|^{2} - |A_{-\nu}^{\mathrm{AM}}|^{2})\nonumber \\
&= \dfrac{1}{6}\nu_{\n}^{\prime}(\epsilon_{\F})\ln\left(1.13\dfrac{\Omega_{\mathrm{BCS}}}{T_{\mathrm{c}}}\right)\sum_{M}M|\gamma_{M}(\rho)|^{2}.
\label{eq-spin-gl}
\end{align}
The magnetization is basically zero when the system is particle-hole symmetric. Equation~\eqref{eq-spin-gl} incorporates the contribution from the slope of the density of states at the Fermi energy. The formula based on the microscopic theory Eq.~\eqref{eq-spin} includes other particle-hole asymmetric contributions such as the one in the gradient expansions in the mixed representation of the order parameter. In order to calculate the spin density and the excitation spectra beyond the first order quantum correction correctly, we need to take account of the feedback effect of the magnetization profile to the order parameter profile.

Figure~\ref{fig-BdG-spin} shows the polarization of spins $\braket{S_{3}(\rho)}$ for several cases with comparison between Eqs.~\eqref{eq-spin} and \eqref{eq-spin-gl}. 
We set an energy cut-off of the summation in Eq.~\eqref{eq-spin} to $15T_{\mathrm{c}}$. 
In the case of $o$ vortices shown in Figs.~\ref{fig-BdG-spin}(a), \ref{fig-BdG-spin}(c), and \ref{fig-BdG-spin}(e), the magnetization at the origin is always zero in the GL picture because the vortex has a normal core [Fig.~\ref{fig-BdG-spin}(a)]. On the other hand, in the microscopic point of view, the quasiparticles with the angular momentum $\ell = 0$ and $-1$ contribute to the spin density at the origin and thus there are finite spontaneous magnetizations as shown by red circles and in the inset of Fig.~\ref{fig-BdG-spin}(a). In Figs.~\ref{fig-BdG-spin}(c) and \ref{fig-BdG-spin}(e), we see finite magnetizations far from the vortex core even for the GL picture. They are attributed to the large difference between $M=2$ and $-2$ components owing to the domain structure. In addition, the magnitudes of these components are large because they are finite in the bulk region. Sign reversals for $\rho \gg \xi_{0}$ are related to the change of the magnitude between $M=2$ and $-2$. The number of sign reversals in Figs.~\ref{fig-BdG-spin}(a), \ref{fig-BdG-spin}(b), and \ref{fig-BdG-spin}(c) are, respectively, 1, 0, and 2. This character is consistent with the order parameter profiles.

In the case of the $v$ vortices shown in Figs.~\ref{fig-BdG-spin}(b), \ref{fig-BdG-spin}(d), and \ref{fig-BdG-spin}(f), finite magnetization occurs at the origin even in the GL picture because there is the finite order parameter as well. However the peak position is shifted in the microscopic point of view by $\rho = b_{\mathrm{c}}$, because an anomalous branch crosses at the finite angular momentum $\ell_{\mathrm{c}} \approx k_{\F}b_{\mathrm{c}}$.

\begin{figure}[t]
\centering
\includegraphics[width=22em]{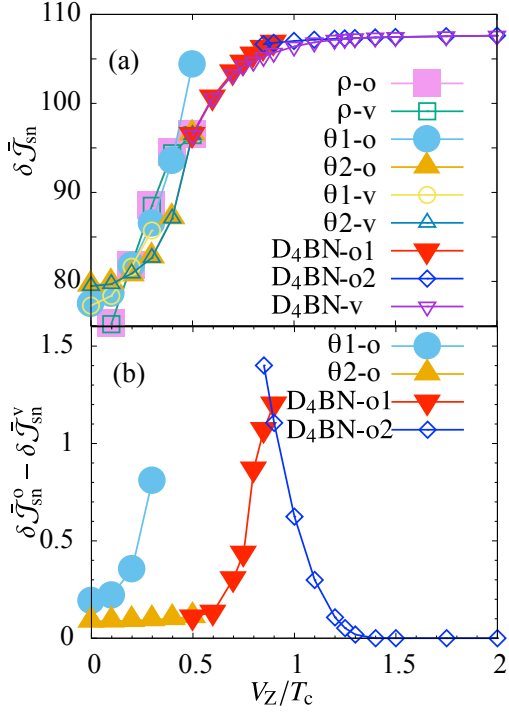}
\caption{Magnetic field dependence of free energy for several vortices. (a) Each free energy is measured from the corresponding uniform state. (b) Free-energy difference of $o$ vortices from corresponding $v$ vortices are shown. At $V_{\mathrm{Z}}/T_{\mathrm{c}} = 0.85$ and $0.9$, the energies of both $D_{4}$-BN-$o$1 and $D_{4}$-BN-$o$2 vortices are measured from that of the unique $v$ vortex.
}
\label{fig-field-dep}
\end{figure}
\subsection{Finite magnetic field}
In this subsection, we work out effects of the magnetic field. In the ${}^{3}$He B-phase, there are several reports on the effects on $v$- and double core vortices on the basis of the GL theory\cite{kasamatsuPRB19,Regan2019}.  In this paper, we restrict ourselves in the axisymmetric cases, and study the field effects on $o$ and $v$ vortices. We show the free energies as functions of the magnetic field in Figs.~\ref{fig-field-dep}(a) and \ref{fig-field-dep}(b). For the magnetic field lower than the critical field between the $D_{2}$-BN and $D_{4}$-BN phases, we distinguish the species of vortices by the direction of the max.~EVs: $\rho$, $\theta$, and $3$. Note that we have seen that two solutions $\theta1$ and $\theta2$ are possible at zero field. The free energies of $\theta2$ vortices become lower than those of $\theta1$ vortices for $V_{\mathrm{Z}}/T_{\mathrm{c}}\gtrsim 0.15$. This may be understood as follows: The magnetic field destabilizes the 3 vortex. An area of the core consisting of the 3 vortex reduces, namely the position of the bump structure goes inside. The loss of the free energy at the bump region decreases because the circumstance along the bump structure becomes small.  On the other hand, the difference in the free-energy density decays slowly for the $\theta1$ vortex, and it is not affected very much by changing the domain position. 

\begin{figure*}[t]
\includegraphics[width = \hsize]{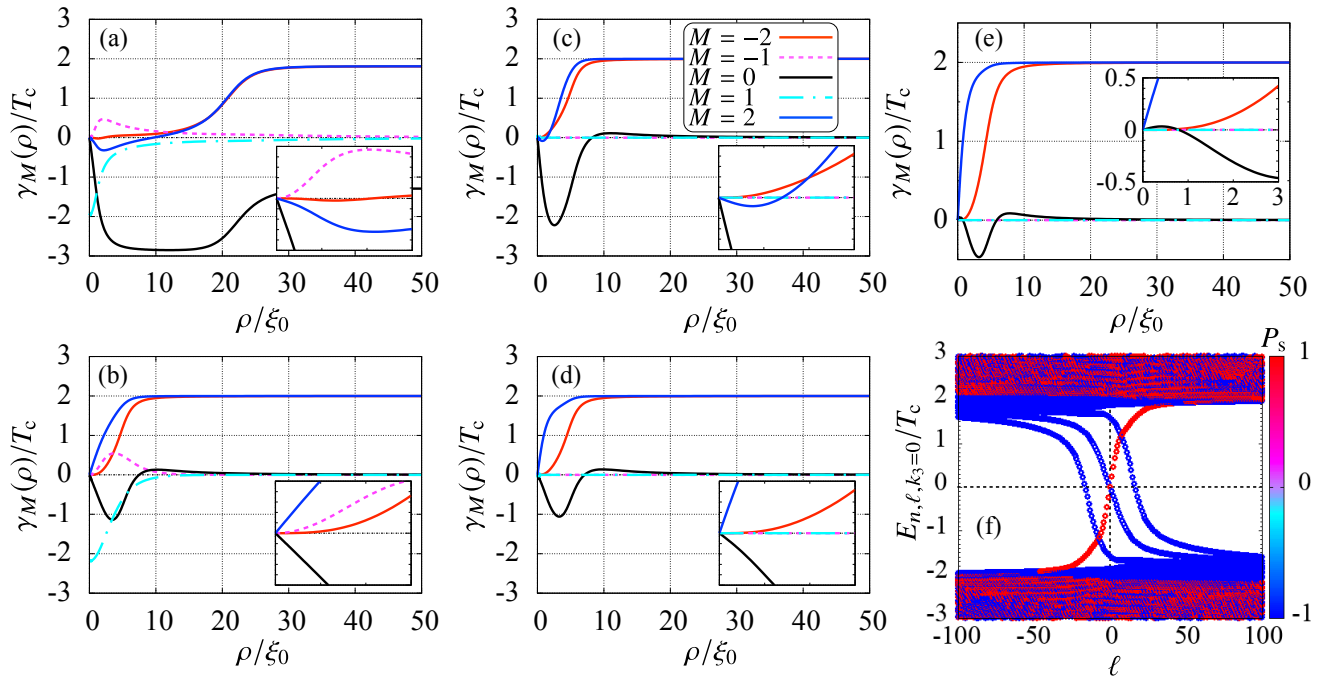}
\caption{
Order parameter profiles of (a) the case at $V_{\mathrm{Z}}/T_{\mathrm{c}} = 0.2$ with the boundary condition schematically shown in  Fig.~\ref{fig-scheme}(b), and (b) -- (d) the cases of $v$-, $o$1- and $o$2 vortices in $D_{4}$-BN phase at $V_{\mathrm{Z}}/T_{\mathrm{c}} = 0.9$, respectively. We also show the order parameter profile and excitation spectra at $V_{\mathrm{Z}} = 1.5T_{\mathrm{c}}$ in panels (e) and (f), respectively. 
The profile in the vicinity of the vortex center is shown in the inset of each panel.
}
\label{fig-rho-D4BN-gap-EV}
\end{figure*}

We can construct $\rho$ vortices for finite $V_{\mathrm{Z}}$, which have domain structures as well as $\theta$ vortices. There are $o$- and $v$ vortices under the axisymmetry. We show the profile of the $v$ vortex at $V_{\mathrm{Z}}=0.2T_{\mathrm{c}}$ in Fig.~\ref{fig-rho-D4BN-gap-EV}(a). It has a domain structure similar to a $\theta2$-$v$ vortex. It should be noted that the gradient energy of a $\rho$ vortex is higher than that of a $\theta$ vortex in contrast to the GL theory. This difference may be attributed to the higher order corrections in the GL expansion. In the $D_{4}$-BN phase, the boundary conditions of $\rho$- and $\theta$ vortices are equivalent, and the branches of $\rho$ and $\theta$ vortices in the state space merge into one branch. Actually Fig.~\ref{fig-field-dep} shows that the free energy of the $\theta2$ vortex becomes the same as that of the $\rho$ vortex at $V_{\mathrm{Z}} = V_{\mathrm{Zc}}$. On the other hand, the $\theta1$-$o$ and $\theta1$-$v$ vortices are destabilized for a decay into $\theta2$-$o$ and $\theta2$-$v$  vortices, respectively, by applying the field. In particular, the $\theta1$-$o$ vortex becomes unstable for $V_{\mathrm{Z}} \gtrsim 0.5 T_{\mathrm{c}}$ and it is different from the $\rho$-$o$ vortex even for $V_{\mathrm{Z}}\ge V_{\mathrm{Zc}}$. We remark that structures of the excitation spectra for $\rho$ vortices are similar to those for $\theta2$ vortices as expected from the similarity in their profiles.

We discuss the field effects in the $D_{4}$-BN phase. The axisymmetry imposes the unique boundary condition except for the global gauge transformation (see Fig.~\ref{fig-scheme}(d)) because the amplitude of the order parameter in the uniform state is isotropic in the momentum space. For $V_{\mathrm{Z}} \lesssim 0.9T_{\mathrm{c}}$, the solutions for $o$- and $v$ vortices can be obtained as ones connected smoothly to the $\rho$ and $\theta2$ vortices.  Their free energies are displayed in Figs.~\ref{fig-field-dep}(a) and \ref{fig-field-dep}(b) by inverted triangles. We find that there are different solutions only for the $o$ vortices at $V_{\mathrm{Z}} = 0.85T_{\mathrm{c}}$ and $0.9T_{\mathrm{c}}$. 
We call these two $o$ vortices in the $D_{4}$-BN phase ``$D_{4}$-BN-$o$1'' and ``$D_{4}$-BN-$o$2'' vortices. The former smoothly connects to $\rho$-$o$ and $\theta2$-$o$ vortices, as shown by the red-inverted triangles in Figs.~\ref{fig-field-dep}(a) and \ref{fig-field-dep}(b), while the latter is shown by blue rhombuses. 
For $V_{\mathrm{Z}} \gtrsim 0.9T_{\mathrm{c}}$, the $D_{4}$-BN-$o$1 vortex is destabilized and the $D_{4}$-BN-$o$2 vortex becomes the unique $o$-vortex solution. 
In Figs.~\ref{fig-rho-D4BN-gap-EV}(b)--\ref{fig-rho-D4BN-gap-EV}(d), we show the solutions of $D_{4}$-BN-$v$, $D_{4}$-BN-$o$1, and $D_{4}$-BN-$o$2 vortices, respectively, at $V_{\mathrm{Z}} = 0.9T_{\mathrm{c}}$. We see slight differences between Figs.~\ref{fig-rho-D4BN-gap-EV}(c) and \ref{fig-rho-D4BN-gap-EV}(d), e.g. the initial slopes of $M=2$ components, but we do not discuss it in details. 
Note that the global $\pi$ gauge transformation is necessary to see a smooth connection between Figs.~\ref{fig-UN-gap}(e) and \ref{fig-rho-D4BN-gap-EV}(c) and that between Figs.~\ref{fig-UN-gap}(f) and \ref{fig-rho-D4BN-gap-EV}(b).

From Fig.~\ref{fig-field-dep}(b), we observe that the free-energy difference between the $o2$- and $v$ vortices decreases after discontinuous transitions from the $D_{4}$-BN-$o1$  to $D_{4}$-BN-$o2$ vortices. Finally the difference vanishes continuously at around $V_{\mathrm{Z}} \sim 1.4T_{\mathrm{c}}$; the $M=\pm1$ components of the $v$ vortices are no longer finite, and the axisymmetric vortex recovers the $P_{1(3)}$ symmetry. The spatial profile of the order parameter at $V_{\mathrm{Z}} = 1.5T_{\mathrm{c}}$ is shown in Fig.~\ref{fig-rho-D4BN-gap-EV}(e). As the Majorana fermions in the $o$-vortex core of the superfluid $^{3}$He B-phase is protected by the $P_{3}$ symmetry, we have confirmed the existence of the Majorana fermion in the $D_{4}$-BN phase within the axisymmetry as shown in Fig.~\ref{fig-rho-D4BN-gap-EV}(f). It should be emphasized that the microscopic calculation of a single vortex in multicomponent superfluids had not been done for finite Zeeman field so far even in the context of the superfluid $^{3}$He-B. We have microscopically demonstrated for the first time that the strong magnetic field actually eliminates $M=\pm1$ components, which break the $P_{3}$ symmetry. 

We give a few comment on axisymmetric vortices in the presence of the magnetic field.
In the case of the superfluid $^{3}$He-B phase, the GL theory suggests that the  nonaxisymmetric double-core vortex is still stable even though the $v$ vortex becomes unstable in the presence of a strong magnetic field. 
We also emphasize that situations may be different in the \tPt superfluids. In this case, the magnetic field affects components which are unaffected in the 
$^{3}$He-B phase, through the property of symmetric tensor, and thereby we expect that the strong magnetic field excludes the possibility of the double core vortex which is stable in the $^{3}$He-B phase. On the other hand, a single vortex in $D_{4}$-BN phase is split into two half-quantized vortices in the GL theory\cite{Masuda:2016vak} using different boundary conditions. It remains important to study a possibility of nonaxisymmetric vortices in \tPt superfluids and the presence of topological zero modes on the basis of microscopic theory.

\section{Summary and discussion}\label{sec:summary}
We have studied axisymmetric vortices in \tPt superfluids, using microscopic theory: the Eilenberger equation to determine the order parameters and the BdG equation to study the eigenenergies and the core magnetization. We have found that several features as a multicomponent superfluid are common to the superfluid $^{3}$He-B phase, though they are overlooked in the GL theory for \tPt superfluids, e.g. the existence of the $v$ vortex. We have shown that the profiles of the core magnetization calculated using the BdG equation are drastically different from those calculated using only the order parameter profiles calculated in the GL theory. 
We have demonstrated that the $o$ vortex is the most stable axisymmetric vortex in the presence of a strong magnetic field, and have found two zero energy Majorana fermions in the $o$-vortex core.
This observation is based on the microscopic calculation in the presence of the magnetic field for multicomponent superfluids and the $P_{3}$ symmetry argument. 

One of open questions is whether two Majorana zero modes in the $o$ vortex give a nontrivial non-Abelian statistics, in contrast to the conventional case of one Majorana fermion zero mode in a vortex resulting in non-Abelian statistics among vortices \cite{Ivanov:2000mjr}, 
which can be generalized to odd numbers of Majorana fermions\cite{Yasui:2010yh,Hirono:2012ad,Eto:2013hoa}.
In the presence of the mirror symmetry, a possibility of the non-Abelian statistics of spin-degenerated Majorana zero modes are considered\cite{Sato2014}. 

The case of a nonaxisymetric case is one of the most important extensions of the present work. 
Although the quadrupole deformation leads to the nonaxisymmetric double-core vortex in the ${}^{3}$He-B phase, the order parameter tensor of \tPt superfluids is restricted to symmetric traceless one and it is nontrivial whether the same deformation may occur or not. 
\footnote{
The effects of the magnetic field on a double-core vortex is not trivial. In the case of ${}^{3}$He-B phase, the $\beta$ phase component does not conflict with the magnetic field\cite{kasamatsuPRB19}. However, the components of the $\beta$ phase and A phase are always the same in \tPt superfluids, and thus the induced components of the double core vortex might be expected to reduce as well as $v$ vortex.
}
Thus we should take account of a general symmetric deformation of an $o$ vortex to a nonaxisymmetric vortex also in terms of the protection of Majorana fermions\cite{Tokuyasu1990,Kobayashi2009a,Klein2014,Silaev2015}. 
It is also important whether 
1/2 quantized non-Abelian vortices in the $D_{4}$-BN phase \cite{Masuda:2016vak} 
admit zero-energy fermions in their cores, and if so whether it may give a novel non-Abelian statistics.

In this paper, we have focused on vortices in the nematic phases, which are the
most stable \tPt superfluid phases in the weak coupling regime under no rotation. 
However, neutron stars are dense neutron matter under extreme conditions, such as rapid rotation. 
Strong coupling effect and rapid rotation might essentially 
change the superfluid phase diagram, and make the cyclic phase and the ferromagnetic phase competitive
to the nematic phases\cite{Mizushima:2016fbn, Mizushima:2017pma}. In particular, it has been found that the cyclic phase admits the 1/3 quantized 
non-Abelian vortices\cite{Semenoff:2006vv,Kobayashi:2008pk}. It is important to explore the microscopic structures of vortices 
and existence of Majorana zero modes in the cyclic and the ferromagnetic phase in \tPt superfluids.

Finally, applications of fermion zero modes to neutron star physics such as contribution to cooling rate and vortex dynamics remain as an important future work\cite{Jones2009}.

\begin{acknowledgments}
The authors thank Michikazu Kobayashi and Shigehiro Yasui for useful discussions. Y.M. thanks Robert~Regan for useful comments.
This work was supported by the Ministry of Education, Culture, Sports, Science and Technology (MEXT)-Supported Program for the Strategic Research Foundation at Private Universities ``Topological Science'' (Grant No.~S1511006). This work was also supported in part by the Japan Society for the Promotion of Science (JSPS) 
Early-Career Scientists Grant No.~JP19K14662 (Y.M.), JSPS Grant-in-Aid for Scientific Research (KAKENHI) [Grants No.~JP16K05448 (T.M.), No.~JP16H03984 (M.N.), and No.~JP18H01217 (M.N.)], and JSPS KAKENHI on Innovative Areas ``Topological Materials Science" (Grant No.~JP15H05855).
\end{acknowledgments}

\appendix
\section{Representation of Cooper pairs} \label{sec:RepCooperPair}
In this Appendix, we summarize the basis for order parameter and matrix representations.
We consider the two basis sets for $A_{\mu j}$. In one basis, $\mu$ and $j$ take $1,\cdots, 3$, and in the other basis they take $+, 0, -$. We distinguish them by calling Cartesian (Car) and Angular Momentum (AM), respectively. We denote them by $A^{\mathrm{Car}}$ and $A^{\mathrm{AM}}$.
Here the relations between two basis sets are given by
\begin{align}
k_{\pm} = \dfrac{k_{1}\pm\ii k_{2}}{\sqrt{2}},~k_{0} = k_{3},\\
\hat{\tau}_{\pm} = \dfrac{\hat{\tau}_{1}\pm \ii \hat{\tau}_{2}}{\sqrt{2}},~\hat{\tau}_{0} = \hat{\tau}_{3}
\end{align}
with $\hat{\tau}_{\mu = 1,2,3} \equiv \ii\hat{\sigma}_{\mu = 1,2,3}\hat{\sigma}_{2}$. 
The matrix for the transformation is given by
\begin{align}
A^{\AM} = U_{\mathrm{T}}U_{4}^{\mathrm{T}} A^{\Car} U_{4} U_{\mathrm{T}}
\end{align}
with
\begin{align}
U_{4} = \dfrac{1}{\sqrt{2}}\begin{pmatrix}
1 & 1 & 0 \\ -\ii & \ii & 0 \\ 0 & 0 & \sqrt{2}
\end{pmatrix} ~~
U_{\mathrm{T}} = \begin{pmatrix}
1 & 0 & 0 \\ 0 & 0 & 1 \\ 0 & 1 & 0
\end{pmatrix}.
\end{align}

We review the decomposition of triplet $p$-wave pairing with respect to the total angular momentum.
Cooper pair has $L=1$ and $S=1$, and thus the  total angular momentum $\bm{J} = \bm{L} + \bm{S}$ takes $J=0,1,2$: $(L=1)\otimes(S=1) = (J=0) \oplus (J=1) \oplus  (J=2)$.
The order parameter tensor $A_{\mu\nu}^{\Car}$, which is defined by $d_{\mu} = A_{\mu\nu}^{\Car}\bar{k}_{\nu}$, satisfies the following properties:
\begin{align}
A \propto 1~\text{for}~J=0,\\ 
A^{\mathrm{T}} = -A~\text{for}~J=1,\\
A^{\mathrm{T}} = A~\text{and}~\Tr A =0~\text{for}~J=2.
\end{align}

Next we see the decomposition of traceless and symmetric tensor, $A^{J=2}$, into the irreducible representations: $\Gamma_{M = -2,\cdots,2}$: $A = \sum_{M} \gamma_{M}\Gamma_{M}$. The basis $\Gamma_{M}$ in Cartesian representation is defined so that 
\begin{align}
(\Gamma_{M})^{*} = (-1)^{M}\Gamma_{-M},~~\Tr \Gamma_{M}\Gamma_{M^{\prime}}^{*} =  \delta_{M,M^{\prime}}.
\end{align}
Therefore we obtain in Cartesian representation 
\begin{align}
\Gamma_{\pm 2}^{\Car} =\dfrac{1}{2}
\begin{pmatrix}
1 & \pm \ii & 0 \\
\pm \ii & -1 & 0\\
0 & 0  & 0
\end{pmatrix},\\
\Gamma_{\pm 1}^{\Car} = \dfrac{1}{2}
\begin{pmatrix}
0 & 0 & \mp 1 \\
0 & 0 & -\ii\\
\mp 1 & -\ii  & 0
\end{pmatrix},\\
\Gamma_{0}^{\Car} = \dfrac{1}{\sqrt{6}}
\begin{pmatrix}
-1 & 0 & 0\\
0 & -1 & 0\\
0 & 0  & 2
\end{pmatrix}.\end{align}
The basis set in terms of the order parameter is explicitly written down, through $\hat{\Delta}_{M}(\bm{k}) = \Gamma_{M,\mu i}\bar{k}_{i}\hat{\tau}_{\mu}$, as
\begin{align}
\hat{\Delta}_{\pm2}(\bm{k}) &= \dfrac{\bar{k}_{1}\pm\ii \bar{k}_{2}}{\sqrt{2}}\dfrac{\hat{\tau}_{1} \pm \ii \hat{\tau}_{2}}{\sqrt{2}} = \bar{k}_{\pm} \hat{\tau}_{\pm}, \\
\hat{\Delta}_{\pm1}(\bm{k} ) &= \dfrac{1}{\sqrt{2}}\left(\mp \bar{k}_{3}\dfrac{\hat{\tau}_{1} \pm\ii \hat{\tau}_{2}}{\sqrt{2}}\mp \dfrac{\bar{k}_{1}\pm\ii \bar{k}_{2}}{\sqrt{2}}\hat{\tau}_{3}\right) \nonumber \\&= \mp\dfrac{1}{\sqrt{2}}\left(\bar{k}_{0}\hat{\tau}_{\pm} + \bar{k}_{\pm}\hat{\tau}_{0}\right),\\
\hat{\Delta}_{0}(\bm{k}) &= \dfrac{1}{\sqrt{6}}\left(-\bar{k}_{1}\hat{\tau}_{1} -\bar{k}_{2}\hat{\tau}_{2} + 2\bar{k}_{3}\hat{\tau}_{3}\right)
\nonumber \\&
= \dfrac{1}{\sqrt{6}}(-\bar{k}_{-}\hat{\tau}_{+} -\bar{k}_{+}\hat{\tau}_{-} + 2\bar{k}_{0}\hat{\tau}_{0}).
\end{align}
The orthonormal basis sets in momentum and spin spaces are given, respectively, by
\begin{align}
\ket{\pm1}_{k} &= \mp \bar{k}_{\pm},~ \ket{0}_{k} =  \bar{k}_{0},\\
\ket{\pm1}_{s} &=  \dfrac{\mp \hat{\tau}_{\pm}}{\sqrt{2}},~ \ket{0}_{s} =  \dfrac{\hat{\tau}_{0}}{\sqrt{2}},
\end{align} 
where $\braket{\cdot|\cdot}_{k}=\frac{3}{4\pi}\int \dd \cos \theta \dd \phi \cdots$, and $\braket{\cdot|\cdot}_{s} = \Tr{\cdots}$. For these notations, $\hat{\Delta}_{M}(\bm{k})$ are represented as
\begin{align}
\hat{\Delta}_{\pm 2} &= \sqrt{2}\ket{\pm1}_{k}\ket{\pm1}_{s} = \sqrt{2}\ket{2,\pm2}_{J},\\
\hat{\Delta}_{\pm1} &= \ket{\pm1}_{k}\ket{0}_{s}+\ket{0}_{k}\ket{\pm1}_{s} = \sqrt{2}\ket{2,\pm1}_{J},\\
\hat{\Delta}_{0} &=\dfrac{\sqrt{2}}{\sqrt{6}}( \ket{1}_{k}\ket{-1}_{s}+\ket{-1}_{k}\ket{1}_{s} + 2\ket{0}_{k}\ket{0}_{s})= \sqrt{2}\ket{2,0}_{J}.
\end{align}

In angular momentum representation, $\Gamma_{M}$ are described as 
\begin{align}
\Gamma_{\pm 2}^{\AM} &=\dfrac{1}{2}
\begin{pmatrix}
1\pm1 & 0 & 0 \\
0 & 0 & 0\\
0 & 0  & 1\mp1
\end{pmatrix},\\
\Gamma_{\pm 1}^{\AM} &= \dfrac{1}{2\sqrt{2}}
\begin{pmatrix}
0 & -1\mp1 & 0 \\
-1\mp1 & 0 & 1\mp1\\
0 & 1\mp1  & 0
\end{pmatrix},\\
\Gamma_{0}^{\AM} &= \dfrac{1}{\sqrt{6}}
\begin{pmatrix}
0 & 0 & -1\\
0 & 2 & 0\\
-1 & 0  & 0
\end{pmatrix}.
\end{align}

\section{Angular momentum}\label{sec:AM}
Here we summarize the definitions of angular momentum operators for Cooper pairs, which are given by
\begin{align}
[L_{\alpha} A(\bm{R})]_{\mu i}^{\Car} &= 
[(L_{\alpha}^{\mathrm{ext}} + L_{\alpha}^{\mathrm{int}}) A(\bm{R})]_{\mu i}^{\Car} \nonumber \\
&= (-\ii \epsilon_{\alpha\beta\gamma}r_{\beta}\partial_{\gamma} A_{\mu i}^{\Car}(\bm{R})+  A_{\mu j}^{\Car}(\bm{R}) \ii \epsilon_{\alpha ji}).
\end{align}
The antisymmetric tensor in the last term can be regarded as the matrix acting on the right subscript of $A$. 
The spin angular momentum operator is given by
\begin{align}
[S_{\alpha} A(\bm{R})]_{\mu i}^{\Car} 
= ( -\ii \epsilon_{\alpha \mu \nu}  A_{\nu i}^{\Car}(\bm{R})).
\end{align}
The total angular momentum is given by $J = L + S = L^{\mathrm{ext}} + J^{\mathrm{int}}$. For the irreducible representations, the $3$rd component of the angular momentum $J_{3}^{\mathrm{int}}$ can be calculated as
\begin{align}
J_{3}^{\mathrm{int}}\Gamma_{M}  = (L_{3}^{\mathrm{int}} + S_{3})\Gamma_{M}  = M \Gamma_{M}.
\end{align}

\section{Rotation of triad}\label{sec:Rotation}
In this Appendix, we discuss rotations of a basis set of the order parameter for intuitive understanding of our boundary conditions.
Let $\widehat{u}$, $\widehat{v}$, $\widehat{w}$ be the triad. Here $\widehat{w} = \widehat{3}$, and $\widehat{u}$, $\widehat{v}$ are obtained by rotation $\widehat{1}$ and $\widehat{2}$ about $\widehat{3}$. Tensors discussed above are described using $\widehat{1}$,~$\widehat{2}$,~$\widehat{3}$. 
For example, in the Cartesian representation, $\Gamma_{M=2,\mu\nu}^{\Car} =( \widehat{1}_{\mu}\widehat{1}_{\nu}- \widehat{2}_{\mu}\widehat{2}_{\nu}+\ii\widehat{1}_{\mu}\widehat{2}_{\nu}+\ii \widehat{2}_{\mu}\widehat{1}_{\nu})/2$. We consider the following triad:
\begin{align}
\widehat{u} &= \cos \varphi \widehat{1} + \sin \varphi \widehat{2},\\
\widehat{v} &=-\sin \varphi \widehat{1} + \cos \varphi \widehat{2}.
\end{align}
Instead, we can write $(\widehat{u},\widehat{v},\widehat{w}) = (\widehat{1},\widehat{2},\widehat{3})R_{\varphi}$, where $R_{\varphi}$ is a rotation matrix around $\widehat{3}$ by angle $\varphi$. Let us summarize this kind of rotation at first. Note that we take the third component of $L$ and $S$ in the same direction $\widehat{3}$.
The rotation operator in real vector (e.g. $\bm{k}$, $\bm{R}$) is performed using
$U_{L}(\varphi) = \exp[-\ii \varphi L_{3}]$. [$U_{L}(\varphi):(\rho,\theta,r_{3})\to(\rho,\theta -\varphi,r_{3})$]. The operation on a function in the Wigner representation $\psi(\bm{k},\bm{R})$ is given by $
U_{L}(\varphi)\psi(\bm{k},\bm{R}) = \psi(R_{\varphi}^{-1}\bm{k},R_{\varphi}^{-1}\bm{R})$.
Therefore, the order parameter is transformed as
\begin{align}
U_{L}(\varphi)d_{\mu}(\bm{k},\bm{R}) &=  d_{\mu}(R_{\varphi}^{-1}\bm{k},R_{\varphi}^{-1}\bm{R}) \nonumber \\
&= \sum_{ij}A_{\mu i}(R_{\varphi}^{-1}\bm{R})R_{\varphi,ji}\bar{k}_{j}.
\end{align}
The simultaneous rotation in real and spin spaces is performed using $U_{J}(\varphi) = \exp(-\ii \varphi J_{3})$ as
\begin{align}
[U_{J}(\varphi)\bm{d}(\bm{k},\bm{R})]_{\mu} &=  \sum_{\nu}R_{\varphi, \mu\nu}d_{\nu}(R_{\varphi}^{-1}\bm{k},R_{\varphi}^{-1}\bm{R}) \nonumber \\&= \sum_{\nu ij}R_{\varphi,\mu\nu}A_{\nu i}(R_{\varphi}^{-1}\bm{R})[R_{\varphi}^{T}]_{ij}\bar{k}_{j}.
\end{align}
Then we can check $R_{\varphi}\Gamma_{M}R_{\varphi}^{T}= e^{-\ii M \varphi}\Gamma_{M}$.

Noting $\hat{1}_{i} = \delta_{1i}$, we see that 
\begin{align}
[\widehat{1}R_{\varphi}^{\mathrm{T}}]_{j} &= \widehat{1}_{i}R_{\varphi,ij}^{\mathrm{T}} = R_{\varphi,1j}^{\mathrm{T}} = R_{\varphi,1i}^{\mathrm{T}}\widehat{j}_{i}
=[\widehat{j}R_{\varphi}]_{1}= \widehat{u}_{j},\\
[R_{\varphi}\widehat{1}]_{j} &= R_{\varphi,ji}\widehat{1}_{i} = R_{\varphi,j1} = \widehat{j}_{i}R_{\varphi,i1}
=[\widehat{j}R_{\varphi}]_{1}= \widehat{u}_{j}.
\end{align}
This reads
\begin{align}
R_{\varphi}\Gamma_{2}^{\Car}R_{\varphi}^{\mathrm{T}} = [\widehat{u}\widehat{u} - \widehat{v}\widehat{v} + \ii (\widehat{u}\widehat{v} + \widehat{v}\widehat{u})]/2.
\end{align}
Therefore the rotation of the triad is described by that  of $\Gamma$.

\begin{figure*}[t]
\includegraphics[width = \hsize]{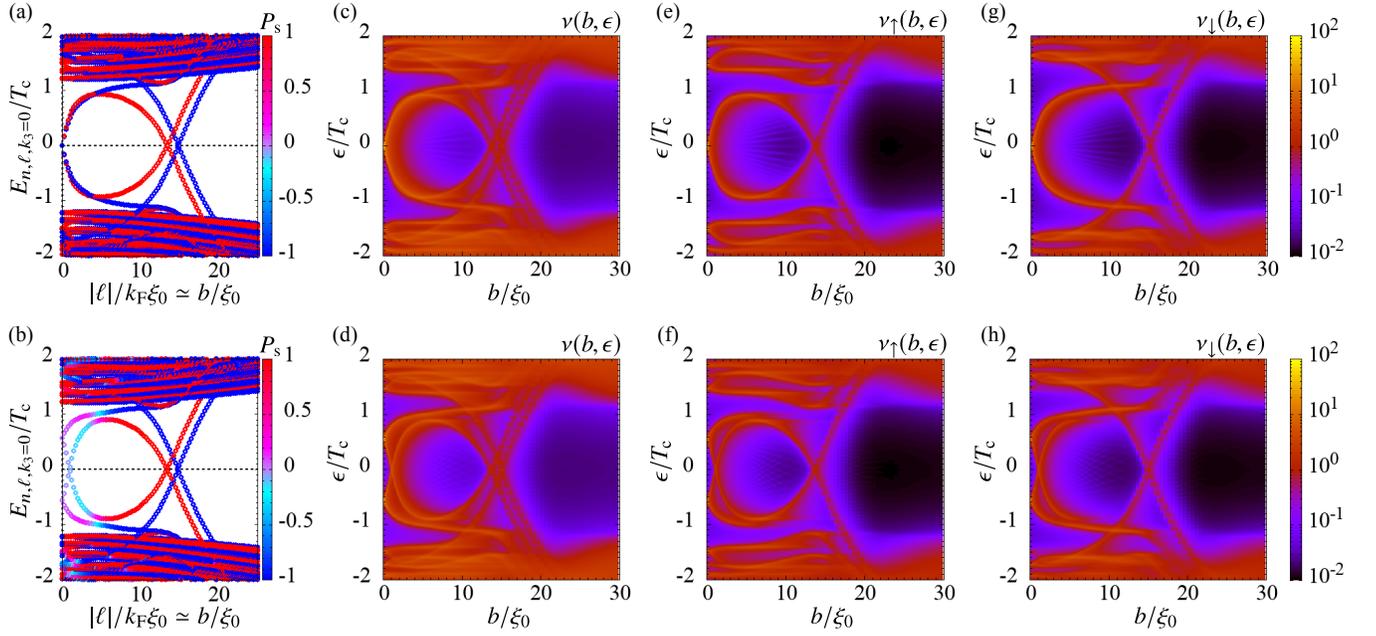}
\caption{
(a), (b) Eigen spectra, (c), (d) LDOS, and (e) -- (f) spin dependent LDOS for $\theta2$-vortices. The top panels are for the $o$ vortex, while the bottom panels are for the $v$ vortex. In the horizontal axes of panels (a) and (b), the angular momentum is converted to the impact parameter $|\ell|/(k_{\F}\xi_{0})\simeq b/\xi_{0}$. 
}
\label{fig-ldos-ev}
\end{figure*}

\section{Comparison between local density of states and eigen spectum}\label{sec:ldos-ev}
The angular momentum of the quasiparticle state is related to the impact parameter in the quasiclassical theory: $|\ell| = k_{\F} b$, where the Bohr--Sommerfeld quantization is necessary for the discreteness of $\ell$. Therefore, the relation of the bound state spectra with the angular momentum is also approximated by the peak positions of the LDOS. The LDOS can be calculated through the retarded quasiclassical Green functions as
\begin{align}
\nu(b,\epsilon) &= \nu_{\uparrow}(b,\epsilon) + \nu_{\downarrow}(b,\epsilon),\label{eq-total-ldos}\\
\nu_{\sigma}(b,\epsilon) &= \int_{0}^{2\pi }\dfrac{\dd \alpha}{2\pi}\mathrm{Re}g_{\sigma\sigma}(\bm{k}_{\F}, b\hat{e}_{1};\ii\omega_{n}\to\epsilon + \ii \eta),\label{eq-spin-ldos}
\end{align}
where $\bm{k}_{\F} = k_{\F}(\cos\alpha\hat{e}_{1} + \sin \alpha\hat{e}_{2})$, and $\eta$ is an infinitesimal positive value set to $0.02T_{\mathrm{c}}$. We only consider $k_{3} = 0$, and compare the LDOS with the eigen spectra obtained by solving the BdG equation for $\theta2$-vortices. The eigen spectra are again shown in Figs.~\ref{fig-ldos-ev}(a) and \ref{fig-ldos-ev}(b) for the $o$ vortex and the $v$ vortex, respectively by changing the horizontal axis to the aboslute value of the angular momentum $|\ell|$ divided by the quasiclassical parameter $k_{\F}\xi_{0} = 4$ for comparison.
In Fig.~\ref{fig-ldos-ev}, the top panels are for the $o$ vortex  and the bottom panels are for the $v$ vortex. We show the total LDOS given by Eq.~\eqref{eq-total-ldos} in Figs.~\ref{fig-ldos-ev}(c) and \ref{fig-ldos-ev}(d). The contribution from each spin component is also given in panels (e)--(f).
In the case of the $o$ vortex, the spin-up and the spin-down sector are not coupled. 
Hence we see the correspondence between the red (blue) symbols in panel (a) and panel (e) (panel (g)). On the other hand, in the case of the $v$ vortex, the spin
sectors are coupled near the vortex core owing to the induced components $\gamma_{\pm1}(\rho)$. 
The effect appears in $P_{\mathrm{s}}$ in Fig.~\ref{fig-ldos-ev}(b). 
Correspondingly, the spin dependent LDOS for the $v$ vortex $\nu_{\uparrow}$ and $\nu_{\downarrow}$ have the intensity on the same energy near the vortex core $b\lesssim 5 \xi_{0}$, while they have different branches for $b\gtrsim 10\xi_{0}$ since $P_{\mathrm{s}} \simeq \pm 1$ in panel (b). After all, for both the $o$ vortex and the $v$ vortex, the consistency between the spectra and the LDOS are confirmed.

\bibliographystyle{apsrev4-1}

\bibliography{library, reference}

\end{document}